# *Likelihood Ratio as Weight of Forensic Evidence: A Closer Look*

**Steven P. Lund and Hari Iyer**

Statistical Engineering Division, Information Technology Laboratory,
National Institute of Standards and Technology,
Gaithersburg, MD 20899

steven.lund@nist.gov
hari@nist.gov

The forensic science community has increasingly sought quantitative methods for conveying the weight of evidence. Experts from many forensic laboratories summarize their findings in terms of a likelihood ratio. Several proponents of this approach have argued that Bayesian reasoning proves it to be normative. We find this likelihood ratio paradigm to be unsupported by arguments of Bayesian decision theory, which applies only to personal decision making and not to the transfer of information from an expert to a separate decision maker. We further argue that decision theory does not exempt the presentation of a likelihood ratio from uncertainty characterization, which is required to assess the fitness for purpose of any transferred quantity. We propose the concept of a lattice of assumptions leading to an uncertainty pyramid as a framework for assessing the uncertainty in an evaluation of a likelihood ratio. We demonstrate the use of these concepts with illustrative examples regarding the refractive index of glass and automated comparison scores for fingerprints.

## Executive Summary

In response to calls from the broader scientific community [1, 2] and concerns of the general public, experts in many disciplines of forensic science have increasingly sought to develop and use objective or quantitative methods to convey the meaning of evidence to others, such as an attorney or members of a jury. Support is growing, especially in Europe [3, 4], for a recommendation that forensic experts communicate their findings using a "likelihood ratio" (see Appendix A for an introduction to likelihood ratios). Proponents of this approach [5-11] appear to believe that it is supported by Bayesian reasoning, a paradigm often viewed as normative (*i.e.*, the *right* way; what someone *should* use) for making decisions when uncertainty exists [12-14].

Individuals following Bayesian reasoning may establish their personal degrees of belief regarding the truth of a claim in the form of odds (*i.e.*, ratio of their probability that the claim is true to their probability that the claim is false) taking into account all information currently available to them. Upon encountering new evidence, individuals quantify their "weight of evidence" as a personal likelihood ratio. Following Bayes rule, individuals multiply their previous (or prior) odds by their respective likelihood ratios to obtain their updated (or posterior) odds reflecting their revised degrees of belief regarding the claim in question. Because the likelihood ratio is subjective and personal, we find that the proposed framework in which a forensic expert provides a likelihood ratio for others to use in Bayes equation is unsupported by Bayesian decision theory, which applies only to personal decision making and not to the transfer of information from an expert to a separate decision maker, such as a juror.

Nevertheless, a likelihood ratio may be viewed as a potential tool for experts in their communications to triers of fact. If a likelihood ratio is reported, however, experts should also provide information to enable triers of fact to assess its fitness for the intended purpose. A primary concern should be the extent to which a reported likelihood ratio value depends on personal choices made during its assessment. Even career statisticians cannot objectively identify one model as authoritatively appropriate for translating data into probabilities, nor can they state what modeling assumptions one should accept. Rather, they may suggest criteria for assessing whether a given model is reasonable. We describe a framework that explores the range of likelihood ratio values attainable by models that satisfy stated criteria for reasonableness. Presenting several such ranges, each corresponding to different criteria, provides the opportunity to better understand the relationship between interpretation, data, and assumptions. We propose the concept of a lattice of assumptions leading to an uncertainty pyramid as a framework for such an analysis.

Recent reports from the U.S. National Research Council and the President's Council of Advisors on Science and Technology [1, 2] primarily focus on the scientific validity of expert testimony, requiring empirically demonstrable error rates. In particular, they promote the value of "black-box" studies [15] in which practitioners from a particular discipline assess constructed control cases where ground truth is known (to researchers, but not the participating practitioners) as surrogates for casework in order to evaluate the collective performance of the discipline. Although we are primarily focused on the use of likelihood ratios, which these reports only tangentially consider, the concerns identified in this article also apply to subjectively selecting the pool of control scenarios required to estimate case-specific error rates. Practitioners adhering to Bayesian principles appear to consider likelihood ratio to be the only logical approach for expert communication, and they seek to implement its use in all forensic disciplines. We acknowledge that likelihood ratios provide a potential tool but emphasize that an extensive uncertainty analysis is critical for assessing when and how likelihood ratios should be used.

In the absence of an uncertainty assessment, likelihood ratios may still be useful as metrics for differentiating between competing claims when adequate empirical information is available to provide some meaning to the quantity offered by the expert. Free of normative claims requiring the use of likelihood ratios, forensic experts may openly consider what communication methods are scientifically valid and most effective for each forensic discipline.

## 1. Introduction

In criminal and civil cases alike, the judicial system involves many individuals making decisions after consideration of some form of evidence (*e.g.*, district attorneys deciding whether or not to file criminal charges, prosecution or defense attorneys deciding or advising their clients whether to accept a plea agreement or proceed to trial, jurors voting guilty or not guilty). These decision makers (DMs) often rely on the findings of forensic experts, whether expressed as a written report or through testimony at a trial, to help inform their decision. How experts express their findings and how DMs factor that information into their ultimate decisions remain areas of great public importance and current research; see, for example, [16-17].

Lindley [18] presented a subjective Bayesian perspective for evaluating the weight of evidence[1] in forensic science.[2] Within this framework, the odds form of the Bayes rule, namely,

$$\text{Posterior Odds}_{\text{DM}} = \text{Prior Odds}_{\text{DM}} \times LR_{\text{DM}} \qquad (1)$$

separates the ultimate degree of doubt a DM feels regarding the guilt of a defendant, as expressed via posterior odds (*i.e.*, probability of guilt after considering the evidence divided by probability of innocence after considering the evidence), into degree of doubt felt before consideration of the evidence at hand (prior odds) and the influence or weight of the newly considered evidence expressed as a likelihood ratio[3] for the DM ($LR_{\text{DM}}$).

In theory, the subjective Bayesian framework provides a uniquely rational and coherent[4] approach for an individual to make decisions in the presence of uncertainty. As such, it has garnered much attention among the statistical forensics community, with many scholars advocating that forensic experts summarize their findings by presenting their own personal *LR* to DMs[5], who could then apply (or envision others applying) Bayes rule to modify their respective prior odds by the reported *LR* and arrive at their posterior odds as to the guilt or innocence of the defendant and choose actions accordingly (*e.g.*, a district attorney decides to file criminal charges, a juror decides to vote not guilty, etc.). This proposed hybrid adaptation can be expressed by the equation

$$\text{Posterior Odds}_{\text{DM}} = \text{Prior Odds}_{\text{DM}} \times LR_{\text{Expert}}. \quad (2)$$

---

[1] The term *weight of evidence* appears in the book *Probability and the Weighing of Evidence* by I. J. Good [19], much earlier than Lindley's *Biometrika* paper [18]. In fact, Chapter 6 in this book is entirely devoted to *weighing of evidence*.
[2] For a general exposure to the potential role of probability and statistics in the law, the reader may consult Fienberg [20], Dawid [21], and Kaye and Freedman [22].
[3] A brief introduction to likelihood ratios is given in Appendix A.
[4] For a systematic introduction to the statistical meanings of "rational" and "coherent," see Lindley [13].

The proclaimed appeal of this hybrid approach is that an impartial expert examiner could determine and convey the meaning of the evidence by computing a likelihood ratio (*LR*), while leaving strictly subjective initial perspectives regarding the guilt or innocence of the defendant to the DM. This adaptation is embraced by many forensic scientists in several European countries and is currently being evaluated as a candidate framework for adoption in the United States. Kadane [23], Lindley [13], and others, however, clearly state that the *LR* in Bayes formula is the personal *LR* of the DM due to the inescapable subjectivity required to assess its value.

Many researchers before us, privately and publicly, have considered whether or not it is appropriate to associate an uncertainty with an *LR* value offered as weight of evidence. The reader may refer to a special issue in Science & Justice [24] that is wholly devoted to this debate. Some of those who adhere to Bayesian decision theory have asserted that it is nonsensical to try to associate an uncertainty to an *LR* since its computation has already taken into account all the evaluator's uncertainty. Others who acknowledge sampling variability, measurement errors, variability in choice of assumptions and choice of models have felt a need to express the effect of such variabilities on an *LR* value by offering an interval estimate (either a Frequentist confidence interval or a Bayesian credible interval) or a posterior distribution.

Our paper explicitly identifies the swap from Equation (1) to Equation (2), and any related claims suggesting that the use of an *LR* to transfer knowledge from an expert to a DM is somehow normative, as having no basis in Bayesian decision theory. We further suggest it is necessary to conduct an uncertainty evaluation regarding the potential difference between $LR_{\text{DM}}$ and $LR_{\text{Expert}}$, requiring consideration of the range of results attainable under a wide-ranging and explicitly defined class of models. This is a broad and systematic view of uncertainty, for which limited sensitivity analyses or use of weighting tools such as Bayesian model averaging will generally be inadequate. Instead, we propose using an assumptions lattice and uncertainty pyramid to enable an audience to evaluate whether an *LR* characterization is fit-for-purpose.

We begin by outlining the general steps required to theoretically evaluate an *LR*.

- The DM constructs a collection of scenarios (*i.e.*, possible sequences of acts of those who may have been involved in the event which is the focus of the legal proceedings or its investigation) to consider. Constructing an *LR* requires partitioning this collection of considered scenarios into two sets. Suppose the DM is a juror who will cast a vote of either 'guilty' or 'not guilty' at the conclusion of a trial. The DM may assign any considered scenario to one of two categories, *guilty* and *not guilty* according to how he or she would vote if that scenario were known to be exactly true. Suppose there are *a* mutually exclusive scenarios under which the DM would declare the defendant to be guilty. For notational convenience we refer to this set as $H_p = \{H_{pi}\}_{i=1}^{a}$. Similarly, we refer to the collection of *b* mutually exclusive scenarios under which the DM would declare the defendant to be not guilty as $H_d = \{H_{dj}\}_{j=1}^{b}$.

- After sorting the set of considered scenarios, the DM assigns his or her (prior) degree of belief in each scenario before considering the totality of trial evidence, *E*. This is done by assigning a probability to each scenario such that the sum of all the probabilities is one. Let the probability assigned to scenarios $H_{pi}$ and $H_{dj}$ be denoted by $\pi_{pi}$ and $\pi_{dj}$, respectively. Denote the sum $\sum_{i=1}^{a} \pi_{pi}$ by $\pi_0$. Then the sum $\sum_{j=1}^{b} \pi_{dj}$ equals $1 - \pi_0$. Here $\pi_0$ is the prior probability from the perspective of the DM that the defendant is *guilty*, and $1 - \pi_0$ is the corresponding prior probability that the defendant is *not guilty*. The conditional probability of scenario $H_{pi}$ given that the defendant is guilty is $w_{pi} = Pr[H_{pi}|H_p] = \dfrac{\pi_{pi}}{\pi_0}$. Similarly, the conditional probability of scenario $H_{dj}$ given that the defendant is not guilty is $w_{dj} = Pr[H_{dj}|H_d] = \dfrac{\pi_{dj}}{1 - \pi_0}$. Note that any scenario not explicitly given a positive prior weight is given a prior weight of zero.[6]

---
[5]*E.g.*, Aitken and Taroni [5] (chapter 3); or the ENFSI guidance document [4], which provides several examples illustrating how forensic examiners may use subjective probabilities to arrive at an *LR* value to convey to the DMs the strength of the evidence they examined. Furthermore, this guidance document also indicates that forensic examiners may convert the numerical *LR* value into a verbal equivalent following some scale of conclusions. Verbal expressions, however, cannot be multiplied by prior odds to obtain posterior odds.

- For each scenario with nonzero prior weight, the DM is to assess the probability of the presented evidence $E$ occurring among all outcomes that could result from the described scenario. Let $L_{pi}$ denote the probability of observing the evidence under scenario $H_{pi}$. (Some may find it more natural to denote this quantity as $Pr[E|H_{pi}]$ but we will use $L_{pi}$ for succinctness.) Similarly, let $L_{dj}$ denote the probability of observing the evidence under scenario $H_{dj}$.

- Once a weight and a likelihood have been determined for each scenario of the observed evidence, the likelihood ratio is given as the sum of the products of the likelihood and the corresponding prior weight for each scenario in the *guilty set* divided by the sum of the products of the likelihood and the corresponding prior weight for each scenario in the *not guilty set*. This may be expressed algebraically as follows:

$$LR = \frac{\sum_{i=1}^{a} L_{pi} w_{pi}}{\sum_{j=1}^{b} L_{dj} w_{dj}}. \tag{3}$$

This formulation highlights that computing an *LR* is generally not free from prior probability assignment at the level of specific scenarios. The *LR* is insensitive to the redistribution of prior weights among scenarios that share a common likelihood within the *guilty set* (or within the *not guilty set*). In the context of source attribution, for instance, the DM may believe the alternative sources are a random sample from a particular population and not have any additional information that would lead to assigning different likelihoods among the alternative sources. In this instance the DM might assign each alternative source a likelihood representing the probability of observing the evidence by random selection from that population, and the denominator becomes that same probability, regardless of what weights $w_{dj}$ would be chosen.

As a more concrete and narrowly focused example, suppose that evidence $y$ has been recovered from a crime scene and that, for simplicity, the DM is only interested in the identity of its source. Further suppose that, given which potential source actually produced $y$, there are no further relevant and unknown details from the perspective of the DM.[7] Let $S_0, S_1, \ldots, S_N$ denote the totality of potential sources, one of which is responsible for $y$. The actual source of $y$ is denoted by $S_q$, where $q$ is unknown. The source $S_0$ is of particular interest to the DM because it is attributed to the defendant. Thus, the primary proposition in question is

$$H_0: \ S_0 \text{ is the source of } y \ (i.e., q = 0).$$

---

[6]A prior weight of zero indicates that the DM would never consider the scenario as plausible regardless of what data were presented. This hardline stance would seem more likely to be taken unintentionally or as a matter of convenience rather than conviction. (Convenience occurring from the fact that the entire collection of scenarios with an assigned prior weight of zero can be removed from further consideration to produce a manageable problem.) Additionally, even the most outlandish scenarios could become seemingly irrefutable, provided sufficient data. By this notion, it seems unlikely that any prior probability is rigid and exactly zero.

[7]*E.g.*, if $y$ were a fingerprint, suppose the only relevant component of uncertainty to the DM is which person, or more specifically which finger, left the impression; or if $y$ consisted of striation marks on a bullet fragment, suppose the DM is only concerned about identifying the gun from which the bullet was fired. For real situations involving multiple pieces of evidence and multiple experts, some forensic scientists suggest the use of *Bayesian Networks*. See, for instance, Taroni, *et al.* [25].

The complement of the proposition $H_0$ is $H_d = H_0^c$, given by

$$H_d : S_1 \text{ or } S_2 \text{ or } \ldots S_N \text{ is the source of } y \text{ (i.e., } q \in \{1, 2, \ldots, N\}).$$

In addition to $y$, suppose one or more control samples (that is, samples from known sources) are available from one or more of the sources $S_j$, $j = 0, \ldots, N$. Denote these, collectively, by $x$.

Suppose $I$ denotes the totality of information available to the DM prior to being exposed to the information supplied by $y$ and $x$. According to the framework Lindley presents, a DM has prior probability $\pi_0 = Pr[H_0|I]$ for the proposition $H_0$ based on whatever information $I$ is available to him or her disjoint from $y$ and $x$. After being informed about the available new information $y$ and $x$, the DM would like to update his or her belief concerning $H_0$ in a *rational* and *coherent* manner.

The DM is interested in $Pr[H_0|y,x,I]$, the probability that $S_0$ is the source of $y$ given all the information available in the crime scene evidence ($y$), the control samples ($x$) and whatever else ($I$). Using the odds form of the Bayes rule, and following Lindley [18], Neumann *et al.* [26] and others, we get

$$LR = \frac{Pr[y|x, H_0, I]}{Pr[y|x, H_d, I]}. \tag{4}$$

In the context of this example, there is only one scenario under which $S_0$ is considered the source of $y$. Hence, the *LR* numerator requires only the conditional probability of $y$ given $x$, $H_0$ and $I$. Suppose this is denoted by $Pr[y|x, H_0]$.[8]

When the number of possible alternative sources is greater than one, evaluating the *LR* denominator, which corresponds to scenarios under which $S_0$ is not the source of $y$, is more complex. The proposition $H_d$ does not say anything about which of $S_1, \ldots, S_N$ is in fact the source. We can decompose $H_d$ as the union of the propositions $H_j$, $j = 1, \ldots, N$, where

$$H_j : S_j \text{ is the source of } y.$$

Because $H_d$ involves multiple scenarios, computing the *LR* denominator requires both a weight and conditional probability of $y$ given $x$ for each $H_j$. Suppose $\pi_0, \pi_1, \ldots, \pi_N$ are the *prior* probabilities, from the perspective of the DM, associated with the propositions $H_j$, $j = 0, 1, \ldots, N$, respectively. Then the denominator of the *LR* takes the form

$$Pr[y|x, H_d] = \sum_{j=1}^{N} w_j Pr[y|x, H_j],$$

where $Pr[y|x, H_j]$ is the probability of $y$ given $x$ and $H_j$ (and $I$) and $w_j = \dfrac{\pi_j}{1 - \pi_0}$. Thus, $w_j$ are the prior probabilities of the DM associated with $H_1, \ldots, H_N$, given $H_0$ is false.

Given the quantities $Pr[y|x, H_j]$, $j = 0, 1, \ldots, N$, and $\pi_0, \pi_1, \ldots, \pi_N$, the *LR* corresponding to $H_0$ is computed as

$$LR = \frac{Pr[y|x, H_0]}{\sum_{j=1}^{N} w_j Pr[y|x, H_j]}.$$

### List of Concerns

The recommendation that an individual substitute someone else's *LR* for his or her own, as represented in Equation (2), is indefensible, rather than normative, under the subjective Bayesian paradigm. Nevertheless, if it can be argued that $LR_{Expert}$ is sufficiently close to $LR_{DM}$, then such a substitution may be acceptable to the DM and fit for his or her purpose. However, there are many reasons why an *LR* value offered by the expert may differ from that of the DM. The following considerations are intended to highlight some of the more prominent subjective choices influencing the value of an *LR*.

---

[8]For simplicity of presentation, we have dropped the term $I$ with the proviso that all probabilities mentioned are *conditional on I*. Furthermore, it is to be understood that expressions such as $Pr[y]$ (or $Pr[y|x]$) refer to marginal (or conditional) probabilities or probability densities depending on whether $y$ is treated as discrete or continuous.

1. **Whose Scenarios?**

   According to the definition of the *LR*, any scenario given a nonzero prior probability by the DM can influence the value of the *LR* and is therefore relevant; scenarios given prior probability zero cannot influence the value of the *LR* regardless of the value of the corresponding likelihood $Pr[y|x,H_j]$, $j = 1,\ldots,N$, and are therefore irrelevant to the DM. Even if $Pr[y|x,H_j]$ is exactly known for any scenario proposed, the *LR* still depends upon the collection of scenarios that are considered as well as the corresponding weights given to them by the DM, neither of which is known to the expert.

   As in the source attribution example above, the set of sources with positive prior probability forms the *relevant population* for the DM. If there are several DMs then each one could have their own set of weights $w_j$ and hence their own relevant population. Given a particular relevant population, the weights assigned to elements of that population can affect the *LR* unless the assigned likelihoods are constant across all members of the population. In particular, $w_j = \frac{1}{N}$ is a special case, not a mandate. How sensitive is the *LR* value to any particular definition of a relevant population?

2. **Whose Likelihoods?**

   In practice, probability functions $Pr[y|x,H_j]$ ($j = 0,1,\ldots,N$) are rarely known in any authoritative sense. A forensic analyst will commonly begin with a prior distribution over a class of models that will then be updated by consideration of empirical data.[9] That is, crime scene data *y* and control data *x* are assumed to be conditionally independent, given the parameter $\theta$ and the event $H_j$, with known distributions $g(y|\theta,H_j)$ and $h(x|\theta,H_j)$ ($j = 0,1,\ldots,N$), respectively. Given $H_j$, $\theta$ is assumed to have a distribution described by the probability function $f(\theta|H_j)$, which is used to express a prior belief about likelihood functions for *x* and *y* given $H_j$ (not to be confused with the prior $\pi_j$, which reflects prior belief in the proposition $H_j$). Hence the joint distribution of *y*, *x*, and $\theta$, given the proposition $H_j$, is described by the probability function

$$a(y,x,\theta|H_j) = g(y|\theta,H_j)h(x|\theta,H_j)f(\theta|H_j). \tag{5}$$

The quantity $Pr[y|x,H_j]$ can be expressed as

$$Pr[y|x,H_j] = \frac{\int a(y,x,\theta|H_j)d\theta}{\int \int a(y,x,\theta|H_j)d\theta dy}. \tag{6}$$

Thus, the distribution of interest for source *j*, $Pr[y|x,H_j]$, has been *exactly* specified through the choice of *f*, *g* and *h*. Asymptotically, as the number of control observations goes to infinity for each potential source $j = 0,1,\ldots,N$, the value of $Pr[y|x,H_j]$ may converge to the same answer for many different choices of *f*, *g* and *h*. In real applications with finite data, however, subjective choices of *f*, *g* and *h* remain influential.

Support for particular choices of *f*, *g* and *h* is sometimes given by showing them to be consistent (as defined by some user-selected process for evaluating such things) with empirical data from similar situations. Even when all DMs agree on what data is appropriate to consider for the case at hand and the criteria to use in assessing whether or not a model is consistent with that data, multiple choices of $f$, $g$, and $h$ will satisfy that requirement. The question remains, how sensitive is the result to any particular modeling choice?

---

[9]This framework includes Bayesian model averaging (BMA) (see Hoeting *et al.,* [27]) whereby the DM specifies a collection of probability model families along with his or her personal probabilities attached to each model. Other DMs implementing BMA may choose differently leading to different model averaging results. Thus, BMA does not remove the need to examine how assumptions affect uncertainty if it is to represent or inform interpretations of multiple individuals. Depending on what is considered to be a *reasonable* class of priors on the model space, the corresponding range of plausible *LR* values may tend to be narrower when using BMA than otherwise.

3. **Approximation**
   When following a subjective Bayesian approach, one uses a definition of *personal* probability that could be viewed as an individual's assessment of a fair value for a bet of $H_0$ versus its complement. It is assumed that for any required probability such a value exists and is unique, and that the individual is able to identify this value without any doubts.[10] Moreover, it is assumed that the collection of specified probabilities satisfy the requirement of *coherence* (*i.e.*, the standard rules of probability are obeyed). Lindley *et al.*, [30] discuss the practical issues one must address in order to reconcile the generally incoherent probability assessments by an individual. They consider several different approaches that one could use in such a reconciliation process. See also Kadane and Winkler [31]. The fact that such reconciliation efforts are necessary points to uncertainties associated with subjective probability assessments. Nevertheless, results derived using such probability models are sometimes treated as free from uncertainties (see, *e.g.*, Taroni, *et al.* [32]).

   Computing an *LR* for anything but the simplest of problems will involve approximations. Rather than assign prior weights that exactly and genuinely reflect one's personal belief, tractable and familiar substitutions are made. In the absence of a rigorous uncertainty analysis demonstrating that the resulting value is sufficiently insensitive to such replacements, the computed value can only provide an approximation of unknown accuracy for the rational and coherent ratio between posterior and prior odds of the DM. Although any DM only needs to be personally satisfied regarding the suitability of using any given *LR* in Bayes formula, guiding the probabilistic interpretation of others requires greater care.

We note that the considerations listed above are not addressed by explaining the assumptions that underlie a given statistical interpretation. Stating assumptions promotes *transparency*, enabling a trained audience to assess whether a presented analysis seems reasonable - much like a statistical hypothesis test. It does not, however, even begin to inform the range of results attainable under alternative analyses that may also be deemed reasonable - the analog of a statistical confidence interval. The *transferability* of an analyst's statistical interpretation (*i.e.*, its value as a surrogate for that of a DM) depends on its robustness[11] across the set of analyses that the DMs would deem plausible.

To assess robustness in a systematic manner, an analyst must first define the space of models to be considered, possibly by providing an explicit *plausibility criterion*, so that robustness has a precise meaning. When extensively characterizing uncertainty, justifying why models in the defined space are reasonable seems less important than justifying why models not in the defined space are unreasonable. The analyst then explores the corresponding range of attainable results by fitting multiple models from within the defined space.[12] In instances where this exploration is incomplete, the full range of plausible results, and thus the suitability of relying on any one particular interpretation, is unknown. To begin to explore the relationship among data, assumptions and interpretations, we consider multiple assumption sets in a form we refer to as the *lattice of assumptions* and present the resulting ranges of *LR*s as an *uncertainty pyramid*. This approach is intended to encourage analysts to explicitly recognize and systematically evaluate the influence of their subjective modeling choices and is illustrated in the following section.

---

[10]Some authors have considered the practical difficulties associated with precisely identifying fair values for bets and this has led to the consideration of *imprecise probabilities*. For a systematic introduction to this topic see, for instance, Walley [28]. This field remains an active area of research (Augustin, *et al.,* [29]).

[11]The book by Morgenthaler & Tukey [33] titled *Configural Polysampling: A Route to Practical Robustness* provides an interesting discussion of the need for considering multiple plausible models and emphasizes the development of robust methods of statistical analysis of data and approaches for assessing small sample robustness of statistical inference procedures.

[12]When a plausibility criterion pertains to a theoretical probability distribution used to model empirical data, the collection of all plausible models defines a region in the space of all cumulative distribution functions (CDFs). This region is sometimes referred to as a *Probability Box* (p-box, for short). See for instance, Williamson [34], Williamson & Downs [35], Hoffman & Hammonds [36], Ferson *et. al* [37], Zhang & Berleant [38-39] Ferson & Siegrist [40], and references contained therein. These and other authors have investigated methods for propagating uncertainties when component distributions are specified in terms of p-boxes.

## The Influence of Modeling Assumptions

We are concerned about seemingly innocuous modeling assumptions latently constraining the space of plausible interpretations as might be presented by a forensic expert. In this section we demonstrate a process for evaluating the restrictive influence of unsubstantiated information that can creep in solely on the basis of distributional assumptions made by an analyst. It should be noted that the data and modeling approaches used in this section are not exhaustive and are not intended to represent analyses generally undertaken by any particular forensic practice. As such, the actual numerical results obtained in this section are not of primary interest. Our intention is to illustrate a process for assessing the influence of modeling assumptions on concrete examples.

Evaluating the influence of a given assumption set (say, assumption set $A$) requires considering the results of multiple analyses, one in which assumption set $A$ was made and others in which different assumption sets (say, assumption sets $B_i, i = 1, 2, \ldots$), *consistent with empirically observed data*, were made. The influence of assumption set $A$ is reflected by the differences among the conclusions drawn upon evaluation of each set of results. In cases where the differences are considered to be substantial, assumption set $A$ has played a critical role, and the conclusion reached from results of the analysis in which assumption set $A$ was made stretches beyond what the data used in the analysis can in fact support. In such a case, it may be inappropriate to rely on any particular assumption set.

### Illustration 1: Glass Example

We consider an educational example discussed in chapter 10 of Aitken and Taroni [5] involving measurements of refractive indices (RI) for glass evidence. Suppose a window is believed to have been broken during the commission of a crime and fragments of glass are recovered from the crime scene. Suppose also that fragments of glass were found on a suspect.[13] Denote by $x_1, \ldots, x_m$ the RIs of the crime scene fragments (bulk sample) and by $y_1, \ldots, y_n$ the RIs of suspect fragments (receptor sample). The two propositions of interest are

$H_p$ : The receptor sample is from the same source as the bulk sample

$H_d$ : The receptor sample is from a different source than the bulk sample.

### Within Source and Between Sources Distributions

Interpreting the information contained in the observed RIs regarding these two propositions requires understanding the distribution of RIs within each source and how that distribution varies from one source to the next. (Note that if the RI distribution did not vary across sources, then the RI observations would not provide any useful information about their source.) Considering how the RI distribution varies from one glass pane to the next results in a distribution of distributions. The collection of possible descriptions or models for the distribution of distributions is overwhelmingly vast. The tendency is to limit the class of potential descriptions by specifying properties of RI distributions that are assumed to remain constant from one window to the next. In particular, the RI distributions across glass panes are often assumed to be identical except for their location (*e.g.*, mean or median). That is, the RI distribution for every potential source is assumed to have exactly the same shape and exactly the same scale (or spread). Such a family of distributions is referred to as a *location family*. This assumption implies that the distribution for the difference between the RIs of each fragment within a glass pane and the median RI of all fragments from that glass pane (*i.e.*, $x - \text{median}(x)$) is exactly the same for any glass pane in the considered relevant population.

---

[13]The fact that fragments of glass were found on defendant's clothes is in itself of evidential value. But, for our illustration we focus only on the source question, as with the illustrative example provided in Aitken and Taroni [5].

In general, the results of analyses (*e.g.*, *LR*) can be highly sensitive to deviations from the assumption that RI distributions differ only by their median from one glass pane to another. Generating empirical confidence in such a strong assumption would require collecting RI data from many windows with enough measurements from each window so as to convince oneself that strictly limiting the set of plausible distributions to a location family will have only a *negligible effect* on the interpretation of the analysis results compared to, for instance, when the shape and scale of the presumed location family are allowed to vary from one source to another. Even with such a vast and consistent dataset, the possibility remains that the RI distribution of any unexamined window differs substantially from the observed characteristics of the other windows. Further illustration of the potential influence of assuming a location family on the interpretation of the observed RI from a particular case is beyond the scope of this paper. That is, the notion of uncertainty we portray in these examples is incomplete. The uncertainty resulting from a more complete examination is expected to be greater than what is illustrated here.

For the sake of simplicity, we proceed by supposing that the informed DM is willing to make the location family assumption. To compute an *LR* for this scenario, let us first introduce some notation. Suppose the cumulative distribution function (*CDF*) of RI values from any single window belongs to the location family of distributions $G(y; \theta) = G_0(y - \theta)$ for some continuous distribution with *CDF* $G_0$ whose median value is zero. Denote the corresponding probability density function (*PDF*) by $g_0$. Furthermore, suppose that, across the (relevant) population of windows, the median RIs $\theta^{(j)}, j = 0, 1, \ldots, N$, are independently and identically distributed (iid) with an unknown *PDF* $f(\theta)$ and corresponding *CDF* $F(\theta)$. That is, we have assumed that $f(\theta|H_j) = f(\theta)$ for all $j = 0, 1, \ldots, N$. For completeness, we display the expression for the resulting *LR* in Equation (7).

$$LR = \frac{\int \left(\prod_{i=1}^{m} g_0(x_i - \theta)\right) \left(\prod_{j=1}^{n} g_0(y_j - \theta)\right) dF(\theta)}{\left(\int \left(\prod_{i=1}^{m} g_0(x_i - \theta)\right) dF(\theta)\right) \left(\int \left(\prod_{j=1}^{n} g_0(y_j - \theta)\right) dF(\theta)\right)}. \quad (7)$$

This example provides an illustration of there being no information available for us to justify assigning different likelihoods to each particular potential source. Hence, we consider the probabilities in the numerator and the denominator of the *LR* from the perspective of a population of windows rather than weighting likelihoods from individual windows according to their prior probability; see related comments following Equation (3).

**Aitken and Taroni Illustrative Analyses**

In the illustrative example provided in Aitken and Taroni [5], it is assumed that $g_0$ is the *PDF* of a normal distribution with a standard deviation equal to 0.00004. That is, the collection of RIs that could be observed from windows are iid according to a normal distribution with unknown window-specific mean $\theta^{(j)}, j = 1, \ldots, N$, and known standard deviation $\sigma$ equal to 0.00004. Lambert and Evett [41], in their Table 10.5, give average RI measurements from 2269 different samples of float glass. Assuming that these measurements are representative of the mean RIs associated with sources $S_j, j = 0, 1, \ldots, N$, Aitken and Taroni apply kernel density estimation, using a Gaussian kernel with varying bandwidths, to estimate the density $f$ (or the *CDF* $F$) from these sample data. The resulting estimates are then used to evaluate the *LR* corresponding to various hypothetical pairs of average RI measurements from the source (window) and receptor (suspect) (see Table 10.6, page 341, Aitken and Taroni [5]). Applying the distribution estimates from Aitken and Taroni to the illustrative example from Evett [42] (see Table 1), and accounting for interval censoring of the recorded measurements to plus or minus 0.00001, produced corresponding *LR*s of 196, 184, and 72, respectively.

Table 1. Refractive Index Measurements from the window and from the suspect

| Measurements from the window | 1.51844 | 1.51848 | 1.51844 | 1.51850 | 1.51840 |
| --- | --- | --- | --- | --- | --- |
| | 1.51848 | 1.51846 | 1.51846 | 1.51844 | 1.51848 |
| Measurements from the suspect | 1.51848 | 1.51850 | 1.51848 | 1.51844 | 1.51846 |

**Multiple Plausible Models**

The consideration of multiple kernel bandwidths for estimating $f$ begins to illustrate the potential uncertainty due to the influence of modeling choices. A more complete evaluation may be obtained by considering how variable the computed *LR* is across the set of all combinations of $g_0$ and $f$ that might be considered plausible. The criteria for establishing the plausibility of a prosed model is personal and likely to vary from one person to the next. However, it is possible for the criteria of a specific individual to be expressed in an objective manner. When criteria for plausibility have been established, the objective intention is to characterize the range of results attainable by any model meeting those criteria[14] rather than to identify a single plausible model (or a narrow set of closely related models in the case of multiple kernel density estimates of $f$ obtained from different bandwidths) and proceeding as though it is the only plausible model or representative of all plausible models.

**Goodness-Of-Fit Tests and Plausibility Criteria**

We note that it is common practice for a data analyst to use a statistical test of goodness-of-fit to assess plausibility of one or more models. In our example, the data modeler could assess the plausibility of a proposed distribution pair ($g_0$ and $f$), given sample data, using any of a number of goodness-of-fit statistical testing procedures. Some well-known methods are: (1) Kolmogorov-Smirnov (KS) test, (2) Cramer-von Mises test, (3) Anderson-Darling test. For related other approaches the interested reader should also consult Owen [43], Frey [44], Liu and Tewfik [45] and Goldman and Kaplan [46]. The concept is the same for each criterion: the data sample itself cannot reduce the space of plausible models to a single *CDF*.

Here we consider the KS test for illustrative purposes. Any other procedure can be used in place of the KS test but the computations can be more challenging. The KS test leads to a confidence band for *CDF*s that are consistent with the data at a prescribed level of confidence, say 95% (see Figure 2 for an illustration of the KS confidence band around the empirical *CDF* for the 2269 RI values reported by Lambert and Evett [41]). When the KS test is used to assess plausibility, any *CDF* that lies entirely within the confidence band would be deemed plausible given the sample data. As the number of observations in the data set increases, the confidence band narrows and the set of plausible distributions is reduced.

**Between-Windows and Within-Window Data Sets for the Glass Example**

We now consider the influence of two data sets on plausible choices for $g_0$ and $f$.

---

[14] Analogous to selecting prior distributions when conducting Bayesian inference, the choice of a plausibility criterion should not be guided by the set of *LR* values it permits, but upon information available before application to the case at hand.

**Table 2.** Refractive index measurements for 2269 glass fragments given in Lambert and Evett (1984).

| RI | Count | RI | Count | RI | Count | RI | Count |
|---|---|---|---|---|---|---|---|
| 1.5081 | 1 | 1.5170 | 65 | 1.5197 | 7 | 1.5230 | 1 |
| 1.5119 | 1 | 1.5171 | 93 | 1.5198 | 1 | 1.5233 | 1 |
| 1.5124 | 1 | 1.5172 | 142 | 1.5199 | 2 | 1.5234 | 1 |
| 1.5128 | 1 | 1.5173 | 145 | 1.5201 | 4 | 1.5237 | 1 |
| 1.5134 | 1 | 1.5174 | 167 | 1.5202 | 2 | 1.5240 | 1 |
| 1.5143 | 1 | 1.5175 | 173 | 1.5203 | 4 | 1.5241 | 1 |
| 1.5146 | 1 | 1.5176 | 128 | 1.5204 | 2 | 1.5242 | 1 |
| 1.5149 | 1 | 1.5177 | 127 | 1.5205 | 3 | 1.5243 | 3 |
| 1.5151 | 1 | 1.5178 | 111 | 1.5206 | 5 | 1.5244 | 1 |
| 1.5152 | 1 | 1.5179 | 81 | 1.5207 | 2 | 1.5246 | 2 |
| 1.5153 | 1 | 1.5180 | 70 | 1.5208 | 3 | 1.5247 | 2 |
| 1.5154 | 3 | 1.5181 | 55 | 1.5209 | 2 | 1.5249 | 1 |
| 1.5155 | 5 | 1.5182 | 40 | 1.5211 | 1 | 1.5250 | 1 |
| 1.5156 | 2 | 1.5183 | 28 | 1.5212 | 1 | 1.5254 | 1 |
| 1.5157 | 1 | 1.5184 | 18 | 1.5213 | 1 | 1.5259 | 1 |
| 1.5158 | 7 | 1.5185 | 15 | 1.5215 | 1 | 1.5265 | 1 |
| 1.5159 | 13 | 1.5186 | 11 | 1.5216 | 3 | 1.5269 | 1 |
| 1.5160 | 6 | 1.5187 | 19 | 1.5217 | 4 | 1.5272 | 2 |
| 1.5161 | 6 | 1.5188 | 33 | 1.5218 | 12 | 1.5274 | 1 |
| 1.5162 | 7 | 1.5189 | 47 | 1.5219 | 21 | 1.5280 | 1 |
| 1.5163 | 6 | 1.5190 | 51 | 1.5220 | 30 | 1.5287 | 2 |
| 1.5164 | 8 | 1.5191 | 64 | 1.5221 | 25 | 1.5288 | 1 |
| 1.5165 | 9 | 1.5192 | 72 | 1.5222 | 28 | 1.5303 | 2 |
| 1.5166 | 16 | 1.5193 | 56 | 1.5223 | 13 | 1.5312 | 1 |
| 1.5167 | 15 | 1.5194 | 30 | 1.5224 | 6 | 1.5322 | 1 |
| 1.5168 | 25 | 1.5195 | 11 | 1.5225 | 3 | 1.5333 | 1 |
| 1.5169 | 49 | 1.5196 | 3 | 1.5226 | 5 | 1.5343 | 1 |

**Table 3.** Refractive index measurements from 49 different locations from a single window. (Data courtesy of Curran (2011))

| | | | | | | |
|---|---|---|---|---|---|---|
| 1.519788 | 1.519901 | 1.519941 | 1.519941 | 1.519941 | 1.519963 | 1.519970 |
| 1.519974 | 1.519974 | 1.519974 | 1.519974 | 1.519974 | 1.519978 | 1.519978 |
| 1.519978 | 1.519981 | 1.519981 | 1.519981 | 1.519981 | 1.519985 | 1.519989 |
| 1.519989 | 1.519992 | 1.519992 | 1.519996 | 1.519996 | 1.519996 | 1.519996 |
| 1.520000 | 1.520000 | 1.520003 | 1.520007 | 1.520007 | 1.520007 | 1.520007 |
| 1.520010 | 1.520010 | 1.520014 | 1.520014 | 1.520014 | 1.520014 | 1.520025 |
| 1.520025 | 1.520029 | 1.520040 | 1.520043 | 1.520047 | 1.520047 | 1.520069 |

**Float Glass Data**

The first data set (See Lambert and Evett, [41]) contains a collection of average RI measurements obtained from various within-window samples collected from different manufactured pieces of float glass. The number of observations contained in each sample is not provided, so sample sizes may vary across the samples and there is some uncertainty as to how this data should be viewed during evidence evaluation. If each sample contained a single observation, the KS confidence band might be used to restrict the marginal distribution of a single RI measurement obtained from a randomly selected window in the population. This marginal distribution is determined by the choice of $g_0$ and $f$ as $h(y) = \int g_0(y - \theta) dF(\theta)$. If the samples consisted only of means of many replicate observations, the KS bounds could serve to restrict the class of plausible choices for $f$, but would not provide much insight for the choice of $g_0$.

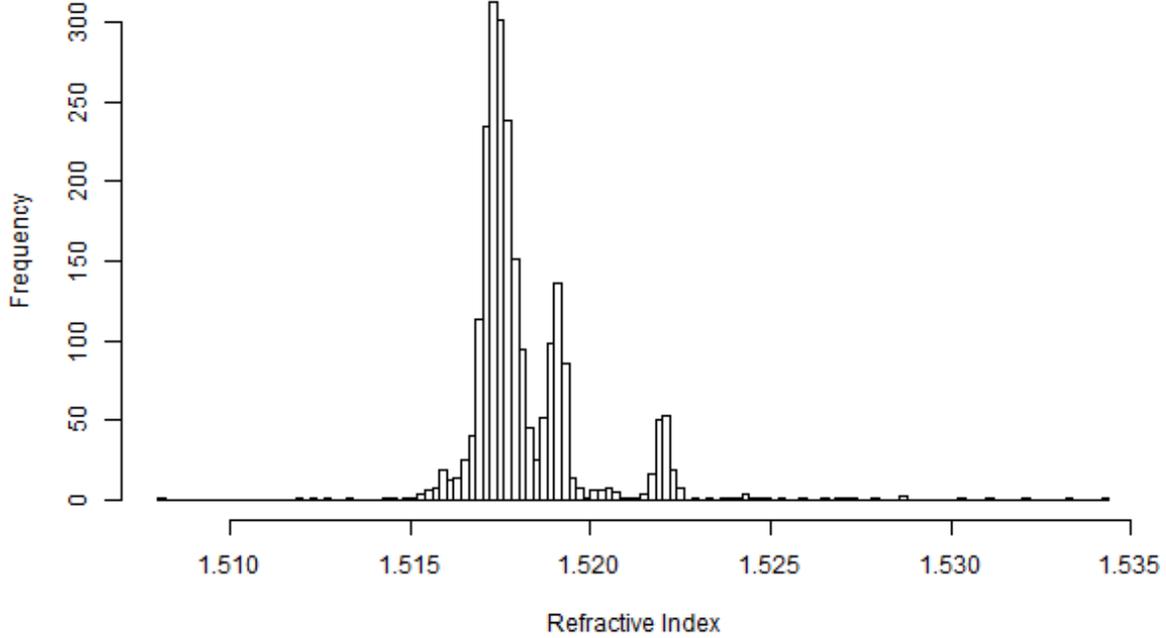

**Figure 1.** Histogram of Glass Data

For illustrative purposes, we treat the data from this set as providing median RI values for a sample of 2269 windows representative of the relevant population. We use the median rather than the mean to reduce the sensitivity of the location parameter $\theta$ to the tails of the distribution $g_0$, which cannot be well-estimated from sample data used in this example. The 2269 reported RI values are shown in Table 2. A histogram of these data is shown in Figure 1. Figure 2 shows the empirical *CDF* (*eCDF*) for these data along with the lower and upper boundaries of a KS 95% confidence band used to define which choices for $f$ will be considered plausible given the *eCDF*. In the lattice of assumptions illustration, we consider several estimates for $f$ based on Gaussian kernel density estimates fit to the 2269 observations with bandwidths spanning from 0 (which corresponds to the *eCDF*) to $2.155 \times 10^{-4}$, which is the maximum bandwidth for which the corresponding discrete distribution obtained by accounting for the reported measurements being interval censored (to plus or minus $1 \times 10^{-4}$) remains entirely within the KS confidence band. *CDF*s for the discrete distributions obtained by accounting for interval censored measurements and the corresponding underlying continuous distributions are shown in Figure 2 for both the *eCDF* and the smoothest kernel density estimate. Kernel density estimates resulting from the intermediate bandwidths of $10^{-5}$, $2 \times 10^{-5}$, $5 \times 10^{-5}$, and $10^{-4}$ are considered during computation but are not displayed. For illustration only, we also include a *CDF* not produced by kernel density estimation. This *CDF*, referred to as Jump, follows the lower KS bound for values less than the mean RI value $m = \dfrac{\sum_{i=1}^{10} y_i + \sum_{j=1}^{5} x_j}{15}$ for the 15 sample fragments, and the upper KS bound for values greater than $m$, with a jump at $m$. This *CDF* is shown in blue in Figure 2. An analyst might feel that the jump distribution is unrealistic and should not be considered. Our point in including it is to emphasize that once a plausibility criterion has been laid down, we must attempt to consider as broad a collection of candidate distributions meeting the criterion as possible; if not, a plausibility criterion and corresponding uncertainty characterization become moving targets.

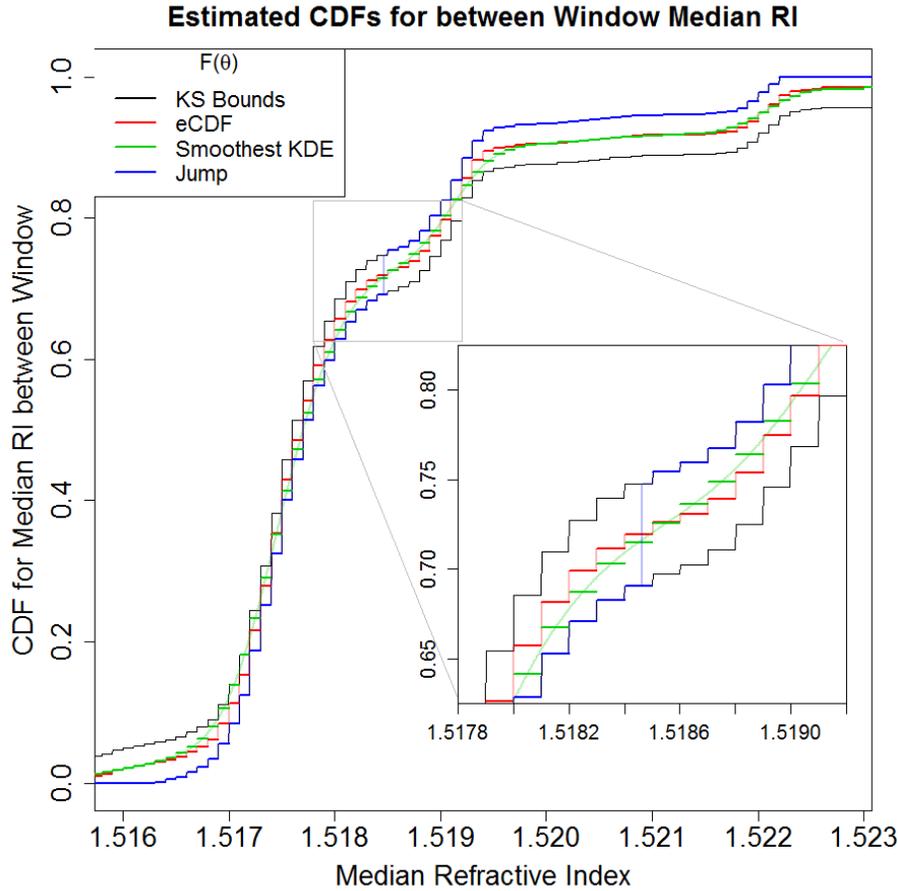

**Figure 2.** 95% Kolmogorov-Smirnov Confidence Band for the Lambert and Evett Glass Data. The bold line segments portray the discrete distribution obtained by accounting for the reported data being interval censored to ± 0.0001. The faded lines display the *CDF* of the underlying continuous distribution.

**Bennett Data**

The second data set consists of 49 refractive index measurements on samples of fragments from 49 different locations on a single window and is used to evaluate the plausibility of within-window distribution choices. These data were collected by Bennett *et al.* [47] and are also mentioned in Curran [48] (see page 42).[15] They are publicly available in the dafs package in R [49]. The original data set consists of RI measurements for a sample of 10 fragments from each of 49 locations on a single window pane for a total of 490 readings. We have selected a single fragment from each of the 49 locations (the listed value in the first row of the bennett.df data frame in dafs). These data are reproduced in Table 3 for the convenience of the reader. For illustrative purposes we treat these 49 RI values as representative of the RI distribution within a single window, providing guidance for choosing $g_0$.

---

[15]Although not explicitly mentioned in Bennett *et al.* [47], these data appear to be interval-censored with variable interval half-widths (approximately) equal to $1.5 \times 10^{-6}$. Consequently, all of our analyses based on these data take this interval-censoring into account.

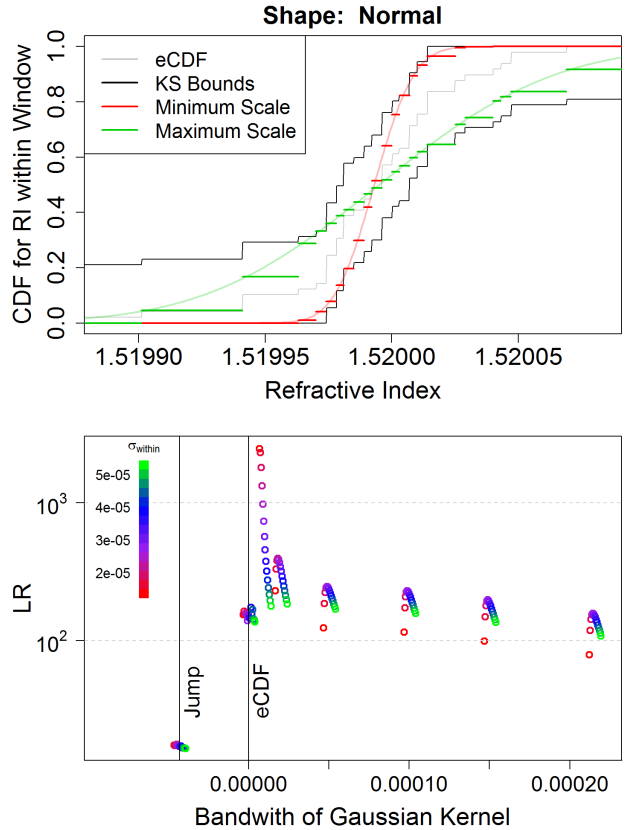

**Figure 3.** Top: 95% Kolmogorov-Smirnov confidence band for the *CDF* of refractive indices from 49 fragments from a single window (Bennett Data). The empirical *CDF* is shown in gray. The faded red and green smooth curves respectively correspond to normal distributions with the smallest and largest scale (standard deviation) parameters such that the discrete distributions obtained, to account for interval-censoring in the reported data (shown using solid red and solid green line segments, respectively), are entirely contained within the confidence band.
Bottom: *LR* values corresponding to various choices of *F*, reflected by position along the x-axis, and the scale factor for the shape corresponding to a normal distribution. The left-most results correspond to the estimate of *F* labeled as Jump, which is displayed in Figure 2. The remaining positions reflect the bandwidth of the Gaussian kernel leading to the estimate of *F* used in computing the *LR*. Within each choice of *F*, the *LR* values are staggered in order of the scale parameter used to define $g_0$ to emphasize the potential non-monotonic relationship between *LR* and scale parameter. The points are color coded to indicate the associated scale parameter values.

For the 49 RI measurements in Table 3, the empirical *CDF* and corresponding KS 95% confidence band are shown in Figure 3. In the lattice of assumptions we consider several distributional shapes including those pertaining to a normal distribution, *t*-distributions with 1 and 0.5 degrees of freedom, respectively, and $\chi^2$ distributions with 2 and 3 degrees of freedom. We also consider a small simulated collection of *nonparametric CDF*s (not belonging to any particular parametric family). Some of these nonparametric *CDF*s fulfill additional constraints of unimodality and/or symmetry. For each considered distributional shape, we identify the range of scale parameters such that the discrete distribution obtained by accounting for the interval-censoring of the reported within-window measurements is contained entirely within the confidence band. For each shape, we consider estimates of $g_0$ obtained at 15 evenly spaced scale values spanning this range. For a given shape and scale parameter, the *LR* is evaluated for each pairing of $g_0$ with each of the choices for $f$ described above. Figure 3 provides a visual summary of the analysis when $g_0$ is assumed to have the shape of a normal distribution. Analogous displays for a subset of other considered shapes are provided in Appendix B.

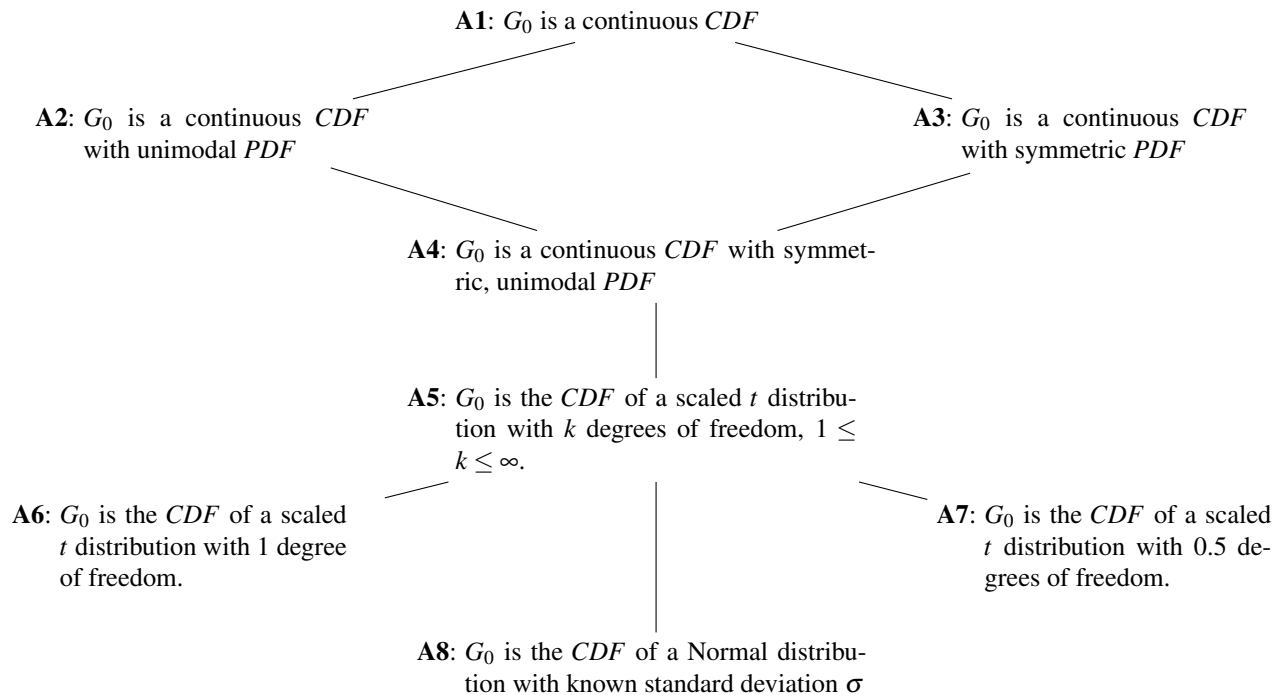

**Figure 4.** Assumptions Lattice for the Glass Example

**Assumptions Lattice**

When modeling the distribution of RI values for fragments from any single window, Lindley [18] assumed normality as do Aitken and Taroni [5]. We recognize this was done for illustrative purposes only. Nevertheless, it is worth noting that normal distributions represent a tiny fraction of *CDF*s meeting the KS criteria, and the impact of exclusively assuming a normal distribution is not clear until the set of *LR* values obtainable by using other distributions lying within the KS bounds have been investigated. In recognition that a given individual's criterion for a distribution to be plausible may include conditions beyond a KS test, in this section we examine the *LR* values obtainable by distributions satisfying a variety of assumption sets. These assumption sets are displayed in the form of a lattice diagram (Grätzer, [50]) as shown in Figure 4. In the figure, when a line segment connects two assumption statements, the assumption appearing lower on the lattice diagram is nested within (*i.e.*, more restrictive than) the assumption appearing higher. In Figure 5, we report interval summaries of the range of *LR* values over the considered subset of the space of all possible models satisfying the criteria for a subset of nodes in Figure 4.

**Discussion of Results**

Results in Figure 5 clearly demonstrate that within this particular educational example the distributional assumptions made regarding the data generating process can have a substantial effect on the *LR* values that would be reported. Keep in mind that we have examined only a small subset of all possible *CDF*s that would be deemed by the KS confidence band to be consistent with the considered RI data. As such, the uncertainty pyramid portrayed in Figure 5 is likely to under-represent the influence of choices of $f$ and $g_0$ within this example. Once again, the point is that reporting a single *LR* value after an examination of available forensic evidence fails to correctly communicate to the DM the information actually contained in the data. Personal choices strongly permeate every model. If expert testimony is to include the computation of an *LR*, we feel an assumptions lattice and corresponding *LR* uncertainty pyramid provide a more honest assessment of the information in the evidence itself and better enable an audience to assess the fitness-for-purpose of the evaluation.[16]

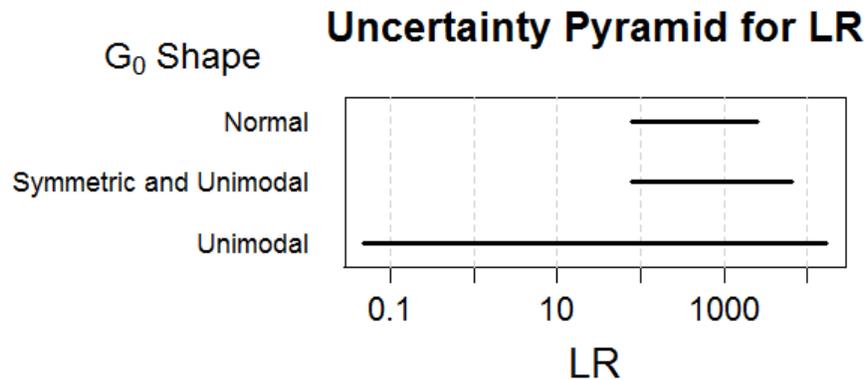

**Figure 5.** Ranges of *LR* values corresponding to subset of choices from the assumptions lattice for $G_0$ combined with Gaussian kernel density estimates for $F$.

**Illustration 2: Score-based Likelihood Ratio based on Simulated Fingerprints**

For this illustration we used a collection of simulated fingerprints to avoid confidentiality issues associated with using real finger marks from actual casework. To be clear, this example does not reflect or assess the behavior or performance of trained latent print examiners. Rather, it is intended to examine the influence of assumptions when forming a score-based likelihood ratio (*SLR*). The ideas expressed in this example are not limited to applications of *SLR*s for friction ridge evaluations, but apply to *SLR* formulations for any comparison discipline.

The software system **Anguli** (see http://dsl.cds.iisc.ac.in/projects/Anguli/index.html; Jadhav, [51]) was used to generate a pair of exemplar-like impressions for 10,000 simulated fingers. One impression from each pair was blurred, occluded, distorted and overlaid on a background image to represent a questioned impression. Minutia were automatically marked in each image using the automatic minutia detecting program MINDTCT (NBIS, [52]) from the National Institute of Standards and Technology (NIST). Figure 6 displays two pairs of simulated images along with the minutia identified by MINDTCT. We retained all detected minutia with a quality score of at least 20. As seen in the Figure 6, this threshold allows erroneous minutia detections; stricter thresholds, however, were found to remove true minutia detections. As the focus here is on the uncertainty in interpreting a given set of scores and not on obtaining the best scores, no formal optimization was performed to select a minutia quality threshold. The BOZORTH3 algorithm (NBIS, [52]) was used to automatically assign a similarity score between two lists of marked minutia.

Suppose a questioned impression (Q) from an unknown source is compared to a test impression ($T_i$) from source $i$ using algorithm $C(\ ,\ )$, resulting in score $s = C(Q, T_i)$. Let $F(s)$ denote the probability of observing score *s* when comparing two images from a common source, and let $G(s)$ denote the probability of observing score *s* when comparing two images from two different sources. Interest lies in the ratio

$$SLR(s) = \frac{F(s)}{G(s)}.$$

---
[16]The proposal to present an uncertainty pyramid is neither intended to replace, nor intended to lessen, the importance of providing objective descriptions of empirical results from analysis and investigation.

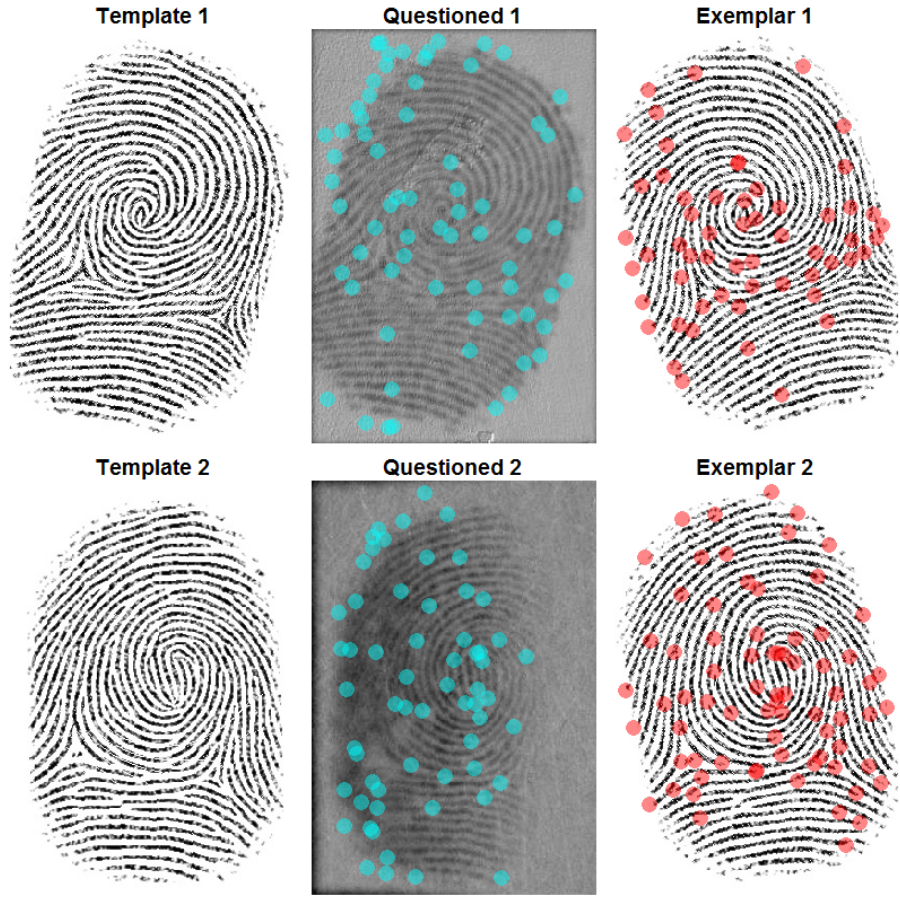

**Figure 6.** Left: Two simulated exemplars used as templates to construct questioned impressions.
Center: Simulated questioned impression. Cyan dots indicate minutia detected by MINDTCT.
Right: Simulated exemplar used for comparison with questioned impressions. Red dots indicate minutia detected by MINDTCT.

To inform possible choices of *F* and *G*, we consider a collection of scores obtained from comparisons for which we know whether or not the compared images originated from a common finger. We refer to comparisons between images generated from the same simulated finger (*e.g.*, comparing Questioned 1 to Exemplar 1 from Figure 6) as "mated." Comparisons between images originating from different simulated fingers (*e.g.*, comparing Questioned 1 to Exemplar 2 from Figure 6) are referred to as "nonmated." Comparing the questioned and exemplar images within each simulated finger produced a collection of $10^4$ mated scores, $S_{M,i}$ ($i = 1, \ldots, 10^4$). We compared questioned and exemplar images independently sampled from their respective collections of $10^4$ images, subject to the constraint that the selected images did not originate from the same simulated finger, to produce a collection of $10^5$ nonmated scores $S_{NM,i}$ ($i = 1, \ldots, 10^5$). The similarity scores output by BOZORTH3 are always nonnegative integers. In our simulation, mated scores ranged from 0 to 227, and nonmated scores from 0 to 36. Scores of 1 and 2 did not occur among any of the mated or nonmated evaluations. The number of occurrences of each integer from 0 to 250 was tabulated for mated and nonmated scores, respectively, and is portrayed in Figure 7. Let $M = [M_0, \ldots, M_{250}]$ and $NM = [NM_0, \ldots, NM_{250}]$ denote the corresponding vectors of occurrences, where $M_j = \sum_{i=1}^{10^4} I_{[S_{M,i}=j]}$ and $NM_j = \sum_{i=1}^{10^5} I_{[S_{NM,i}=j]}$. Here, the term $I_{[S=j]}$ is equal to 1 when $S = j$ and 0 otherwise.

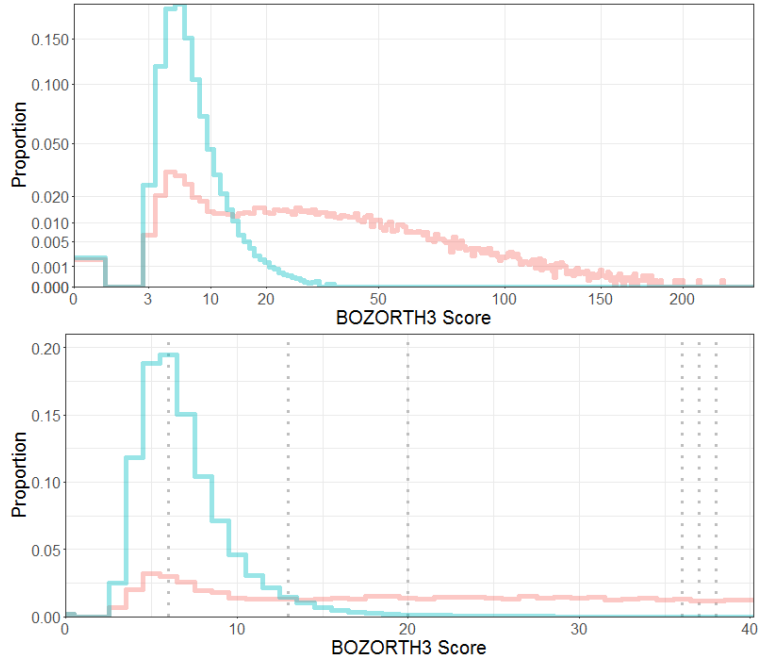

**Figure 7.** Top: Histogram reflecting proportion of simulations resulting in each score for the mated (red) and nonmated (blue) pairs. Note that *x* and *y* axes are provided on a square root scale.
Bottom: A zoomed-in view, using linear scales. The dotted lines indicate the scores for which the *SLR* range was examined.

The choice of reference comparisons that are suitable for informing the score distributions *F* and *G* for a particular case is subjective and influential. In this illustration, we ignore this choice as a potential source of uncertainty and operate as though all DMs have agreed on the simulated collection of scores as being exclusively appropriate for informing their beliefs. That is, we suppose that all relevant DMs consider the mated scores to be iid from *F* (i.e. $S_{M,i} \sim F$) and the nonmated scores to be iid from *G* (i.e. $S_{NM,i} \sim G$).

The *SLR* value corresponding to the score observed for a particular comparison varies as one considers various plausible sets of assumptions used to evaluate *F* and *G*. In this exercise, we examine *SLR* ranges for scores of 6, 13, 20, 36, 37, and 38. The scores 6, 13, and 20 were chosen because the corresponding ratio of relative frequencies were near 0.1, 1, and 10, respectively. The scores 36, 37, and 38 were chosen to examine the robustness of the *SLR* at and just beyond the most extreme observed nonmated score (36 in our illustration).

We consider a different plausibility criterion here than was used in the glass example. Suppose the proposed mated probability mass function (*PMF*) is given by $\boldsymbol{F}' = [F'_0, \ldots, F'_{250}]$, where $F'_j = Pr(S_M = j)$. Similarly, let the proposed nonmated *PMF* be given by $\boldsymbol{G}' = [G'_0, \ldots, G'_{250}]$, where $G'_j = Pr(S_{NM} = j)$. We consider the test statistic

$$Z_{\boldsymbol{F}',\boldsymbol{G}'} = \sum_{i=0}^{250} \left[ \frac{(E_{M,i} - M_i)^2}{E_{M,i}} + \frac{(E_{NM,i} - NM_i)^2}{E_{NM,i}} \right], \tag{8}$$

where $E_{M,i} = F'_i \times 10^4$ and $E_{NM,i} = G'_i \times 10^5$ are the expected counts associated with a score of *i* under $\boldsymbol{F}'$ and $\boldsymbol{G}'$, respectively. The tables of observed counts include many cells with small values, so we estimate the sampling distribution of this test statistic under proposed distributions $F'$ and $G'$ using simulation rather than relying on an asymptotic chi-squared approximation. That is, in each of many iterations, we draw $M^* \sim \text{multinomial}(10^4, F')$ and $NM^* \sim \text{multinomial}(10^5, G')$ and use the simulated values to obtain $Z^*_{F',G'}$, computed from Equation (8) with $M^*$ and $NM^*$ in place of $M$ and $NM$, respectively. The collection of $Z^*_{F',G'}$ values is used to asses whether $Z_{F',G'}$ is lower than the $95^{th}$ percentile of the test statistic in Equation (8) under the null distribution where $F'$ and $G'$ are exactly correct. If so, then $F'$ and $G'$ are considered plausible.

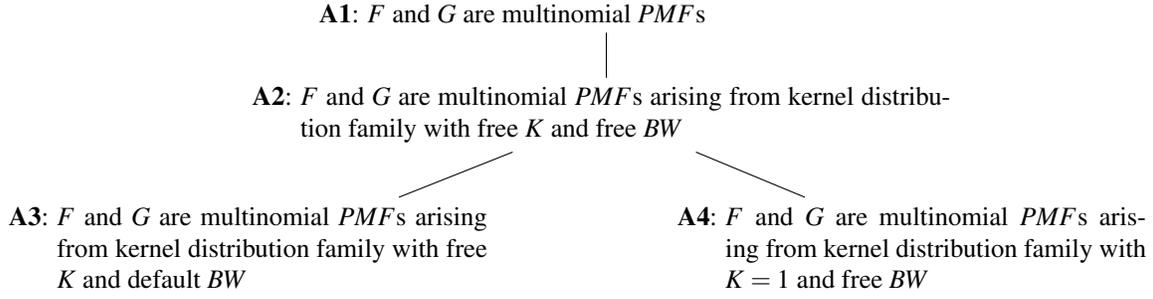

**Figure 8.** Assumptions Lattice for the Fingerprint Example

We evaluate the range of *SLR* values attainable from distributions meeting the criteria described above, first while considering any $F'$ and $G'$ as candidates and then considering only those belonging to various classes of Gaussian kernel distribution estimates applied to power transformations of the observed scores. More precisely, we consider kernel distribution estimates of the form

$$F'_{K_1, BW_1}(s) = 10^{-4} \sum_{i=3}^{227} M_i \times \frac{\Phi\left(\frac{(s+0.5)^{K_1} - i^{K_1}}{BW_1}\right) - \Phi\left(\frac{(s-0.5)^{K_1} - i^{K_1}}{BW_1}\right)}{1 - \Phi\left(\frac{2.5^{K_1} - i^{K_1}}{BW_1}\right)}$$

and

$$G'_{K_2, BW_2}(s) = 10^{-5} \sum_{i=3}^{36} NM_i \times \frac{\Phi\left(\frac{(s+0.5)^{K_2} - i^{K_2}}{BW_1}\right) - \Phi\left(\frac{(s-0.5)^{K_2} - i^{K_2}}{BW_2}\right)}{1 - \Phi\left(\frac{2.5^{K_2} - i^{K_2}}{BW_2}\right)}$$

for $s \geq 3$, where $0 < K_1, K_2 \leq 1$; $BW_1, BW_2 \geq 0$; and $\Phi(\cdot)$ denotes the *CDF* of a standard normal distribution. For completeness, define $F'_{K_1, BW_1}(0) = M_0 \times 10^{-4}$, $F'_{K_1, BW_1}(1) = 0$, and $F'_{K_1, BW_1}(2) = 0$. Similarly, define $G'_{K_2, BW_2}(0) = NM_0 \times 10^{-5}$, $G'_{K_2, BW_2}(1) = 0$, and $G'_{K_2, BW_2}(2) = 0$. Note $BW_1 = 0$ corresponds to $F'$ being the empirical *PMF* for the mated scores, and $BW_2 = 0$ corresponds to $G'$ being the empirical *PMF* for the nonmated scores. We also consider the class of distributions where $K_1$ and $K_2$ are fixed at 1 (still allowing $BW_1, BW_2 \geq 0$), and the class of distributions where $BW_1$ and $BW_2$ are the bandwidth selections produced by applying the R function `density` (R Core Team, 2017) with default settings to the sets $\{M_i^{k_1}\}_{i=1}^{10^4}$ and $\{NM_i^{k_2}\}_{i=1}^{10^5}$, respectively (allowing $0 < K_1, K_2 \leq 1$). The distributions produced using $k_1 = k_2 = 1$ and default bandwidths did not pass the plausibility criterion as the corresponding value of $Z_{F',G'}$ was near the $99^{th}$ percentile of the null distribution. The assumptions lattice for the considered classes of distributions is shown in Figure 8. Corresponding *SLR* ranges are presented as uncertainty pyramids in Figure 9.

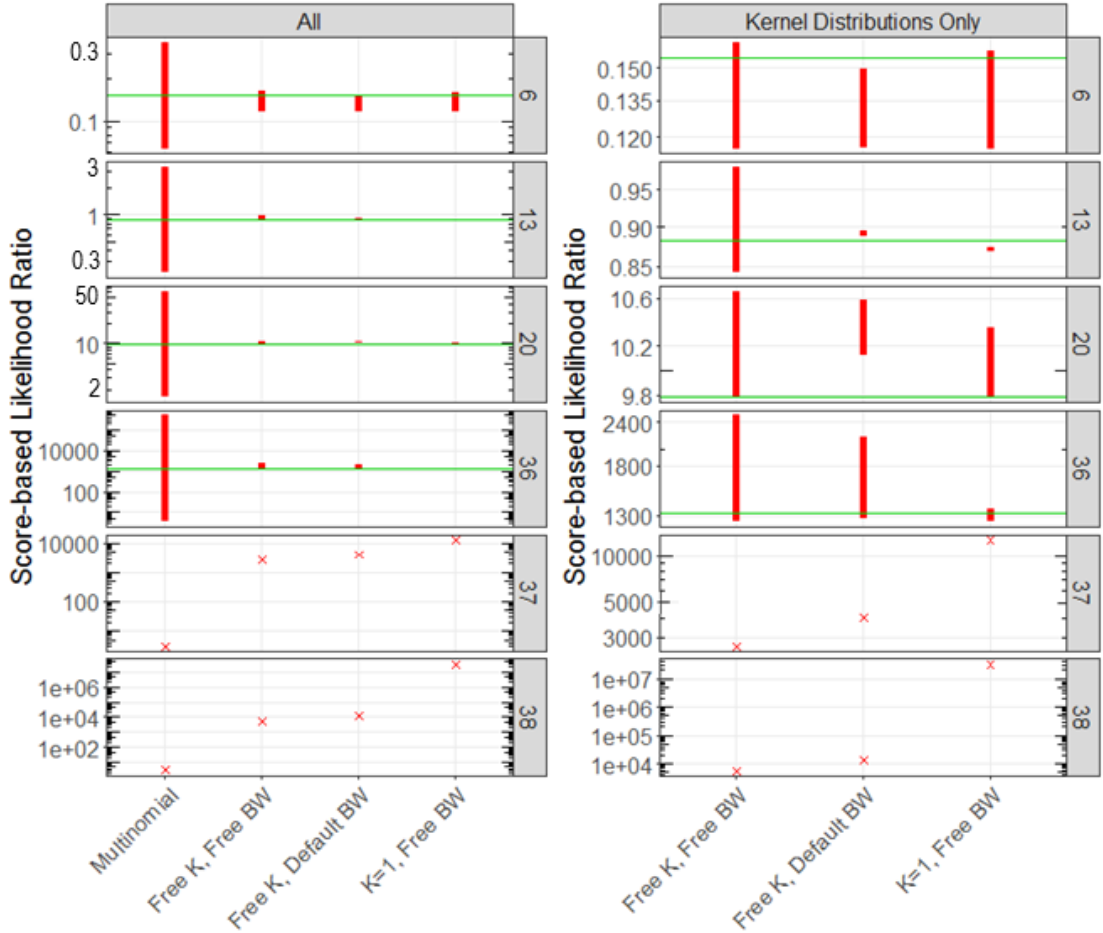

**Figure 9.** *SLR* uncertainty pyramids for various scores. Panels are vertically arranged according to the score for which the *SLR* is computed. The right panel excludes results from the general multinomial class in order to better depict the results from the classes of kernel distribution estimates. Green horizontal lines depict the ratio of relative frequencies, $10 \times \frac{M_s}{NM_s}$. Vertical line segments depict the range of *SLR* attainable by distributions satisfying the selected plausibility criterion and belonging to class indicated along the $x$-axis. Points shown in bottom two panels corresponding to scores of 37 and 38 indicate the lower bound of the *SLR* range. The corresponding upper limits and ratio of relative frequencies are all positive infinity.

## Discussion

The viewpoints expressed in this paper are largely motivated by considerations of standard practices in measurement science, a discipline for which a fundamental purpose is to facilitate meaningful communication regarding properties of an object or system among interested parties. From the perspective of metrology, the hybrid *LR* framework asks a forensic expert to measure the weight of evidence on behalf of the DM and report its value for subsequent use in Bayes formula. As a measurement, any provided *LR* value would require an accompanying uncertainty statement (JCGM, [53]; Possolo, [54]) characterizing the analyst's belief regarding its deviation from the "true value," which the Bayesian paradigm defines as the *LR* value a given DM would arrive at following careful review of the complete body of evidence considered by the expert. Overlooking or dismissing the relevant uncertainty would treat the value obtained by an expert as though it is a perfect measurement of weight of evidence, universally and exactly accurate. This directly contradicts the Bayesian paradigm where no such value can be assumed to exist, as the *LR* is a personal and subjective entity.

Although our discussion of the *LR* has centered around the perspective of Bayesian decision theory, our concerns apply to any framework motivating the use of an *LR* as a means for experts to communicate their findings. Whether a probability is intended to be personal or communal, it is not empirical in the sense that it is not directly observable. A model is required in order to translate data into a probability, and the question of how robust the translation is among reasonable model choices remains central.

We do not make a recommendation regarding when an uncertainty characterization yields a particular *LR* result as being fit for purpose. Our hope is that policy makers will assess the adequacy of relying on *LR* characterizations in the context of the framework presented here, mindful of what range of alternative results might be reasonably attained and of the criteria used to make that assessment. One might expect to find the least degree of uncertainty in applications of probabilistic evaluation of high-template, low-contributor DNA samples, and we recognize that the community may be well-founded in its use of probability to facilitate knowledge transfer in such cases. We do not view this as an exception to framework we present, but rather as a scenario in which extensive uncertainty evaluations would likely yield a degree of consensus leading most people to conclude an offered *LR* value is fit-for-purpose. Forming a lattice of assumptions and uncertainty pyramid, including explicitly identifying what data will be considered, for applications in the field of high-template, low-contributor DNA evaluations could help provide clarity to other forensic disciplines seeking to demonstrate or develop a basis for using a similar *LR* framework. In absence of a suitable uncertainty characterization, or when the uncertainty is deemed too large, *LR* values may require less literal interpretations.

When an *LR* value is the output of a computer algorithm, one may reasonably assume that, given the inputs, it is highly reproducible. In this sense, an *LR* value may be transferable as a discriminant score rather than the ratio of two probabilities. In this context, a discriminant score attempts to produce an optimal ordering among a collection of independent scenarios that may originate from either $H_p$ or $H_d$. For a given ordering, a decision rule is indicated by a threshold, with all scenarios having a score to one side of the threshold being ascribed to $H_d$ and scenarios with scores on the other ascribed to $H_p$. The ordering is optimal when any chosen threshold corresponds to minimizing the error rates given the total number of scenarios that will be ascribed to $H_p$ and $H_d$. In a theoretical scenario where the true *LR* is known for each scenario, the *LR* is the optimum discriminant score. When viewed as a discriminant score, an *LR* value would not have direct, probabilistic interpretation, as its meaning only becomes apparent from its relative positioning to *LR* determinations for other scenarios, evaluated by the same process, including suitable, controlled reference applications. The effectiveness of a given scoring method can be empirically assessed using Receiver operating characteristic (ROC) plots (Peterson and Birdsall, [55]; Green and Swets, [56-57]).

Relying on a given scoring method, an expert could provide demonstrations or scientifically sound descriptions to answer many helpful questions. For instance, in a source-level evaluation, an expert might address:

- How are scores produced and why? What collection of reference scenarios are used to evaluate the performance of the considered scoring methods? How are these chosen in light of the considered case?

- What score was obtained corresponding to the source of interest?

- What alternative sources were considered and what were the corresponding scores?

- How do the scores from this particular case compare to the scores obtained among the reference collection used to evaluate method performance?

More broadly, objective descriptions of procedures followed and outcomes obtained throughout investigation of the case and broader experience may present a promising path to ensuring transferability of information from a forensic expert to DMs.

## Summary

The *LR* framework has been portrayed by some as having an exclusive, normative role in forensic expert communication on the basis of arguments centered around mathematical definitions of rationality and coherence (*e.g.*, Biedermann *e*t al., [58]). These arguments are aimed at ensuring a form of self-consistency of a single, autonomous decision maker.[17] Decision theory, however, does not consider the transfer of information among multiple parties as occurs throughout the judicial process when one or more DMs rely on forensic experts to help inform their decisions. Thus, while decision theory may have a normative role in how a DM processes information presented during a case or trial in accordance with his or her own personal beliefs and preferences, it does not dictate that a forensic expert should communicate information to be considered in the form of an *LR*.[18]

Bayesian decision theory neither mandates nor proves appropriate the acceptance of a subjective interpretation of another[19], regardless of training, expertise or common practice. It does not recognize one person's subjective inputs as superior to those of another, and therefore does not support any one particular *LR* value. Validation efforts can demonstrate that the interpretation corresponding to a particular model is reasonable, but should not be misunderstood to mean the model is accurate or authoritatively appropriate.[20]

Validation efforts can also inspire an explicit plausibility criterion. By conducting multiple analyses attempting to span the space of assumptions meeting a specified plausibility criterion, an analyst can purposefully explore the robustness of an interpretation. Presenting an uncertainty pyramid, along with an explanation of the corresponding plausibility criterion and a description of the data, may provide the audience the opportunity for greater understanding of the interactions among data, assumptions and interpretation. The audience may then, more reasonably, assess whether any particular result is fit-for-purpose.

If such uncertainty characterizations are considered untenable for a given application, one may be forced to conclude that the *hybrid plan* (see Equation (2)), though appealing, is impractical to implement. It does not mean that, just because one is unable to calculate the required value, one should accept the value that can be calculated.

We hope this paper will encourage the forensic science community to be mindful of the many subjective components involved in any interpretation. Correspondingly, we hope best-practice guidances address how to avoid overstating the authority or rigor underlying any particular interpretation of evidence and require a presentation of uncertainty. Additionally, we hope the forensic science community comes to view the *LR* as one *possible*, not normative or necessarily optimum, tool for communicating to DMs. We hope such viewpoints will increase the priority given to developing tools for descriptive presentations that meet the strict standards of scientific validity by focusing on empirical or reproducible results, assisting the DMs in directly establishing their own respective interpretations of evidence.

## Acknowledgments:

The authors are grateful for the valuable feedback received from the reviewers of this paper. In particular, we would like to acknowledge William Guthrie, Dr. Martin Herman, Dr. Adam Pintar, Professor David Kaye, Professor Karen Kafadar, Professor J. Kadane, Professor Hal Stern, Dr. John Butler, Dr. Jonathon Phillips, Dr. Antonio Possolo, and Professor Jacqueline Speir for their detailed comments and suggestions, which were very helpful in making a number of substantial improvements to the manuscript. SPL would also like to acknowledge Jessica Lund for her tireless efforts that made this paper possible.

---

[17] More specifically, the Dutch book arguments (Hájek, [59])

[18] Some may argue that because any given DM is likely unfamiliar with formal decision theory, a trained expert should act on their behalf to form an *LR*. As expounded throughout this paper, the interpretation of evidence in the form of an *LR* is personal and subjective. We have not encountered any basis for the presumption that the surrogate *LR* of an expert will reflect a truer implementation of decision theory than will the unquantified perception of the DM following effective presentation of the information upon which the expert's *LR* is based.

[19] "I emphasize that the answers you give to the questions I ask you about your uncertainty are yours alone, and need not be the same as what someone else would say, even someone with the same information as you have, and facing the same decisions." – Kadane [23]

[20] As a result of this common misunderstanding, we prefer phrases that use "plausible" in place of "validated."

# Appendix A: Likelihood Ratio Introduction

The concept of likelihood ratio (*LR*) arises naturally when one is faced with the problem of deciding whether an observation $x$ came from one of two populations. Consider a simple situation involving two urns, urn 1 and urn 2. Urn 1 has 99 red balls and one green ball, and urn 2 has 99 green balls and one red ball. One of the urns is chosen (we do not know which one or the process used to make the choice) and, after thoroughly mixing the balls in it, one ball is selected and its color is noted. Suppose the ball is red. We would like to know whether the ball is from urn 1 or urn 2.

One may proceed as follows. Let us assume that every ball from the chosen urn had an equal chance of being chosen. Then, if urn 1 was chosen, the probability of drawing a red ball is 99%. If urn 2 was chosen then the probability of drawing a red ball is 1%. Thus, a red ball is 99 times more likely to be drawn if urn 1 was chosen than if urn 2 was chosen. That is, the ratio

$$\frac{\text{Probability of drawing a red ball given urn 1 was chosen}}{\text{Probability of drawing a red ball given urn 2 was chosen}} = 99. \quad (A.1)$$

Whatever the initial belief might have been of an individual regarding where urn 1 or 2 was selected, the effect of the observing a red ball is likely to encourage the individual to update their beliefs by increasing the probability they initially assigned to the scenario that urn 1 was selected.

The above example provides the beginnings of the concept of weight of evidence. It also suggests that the ratio of probabilities of an observed occurrence under each each of the two considered scenarios must play a role in adjusting one's prior beliefs regarding which scenario is true. The ratio in equation (A.1) is called the likelihood ratio for urn 1 corresponding to the observation of a red ball. More generally, if $x$ denotes data observed from one of two distributions, $f_1$ or $f_2$, then the ratio

$$\frac{\text{Probability of observing } x \text{ given } x \text{ came from } f_1}{\text{Probability of observing } x \text{ given } x \text{ came from } f_2}$$

is called the likelihood ratio for $f_1$ corresponding to the observation $x$. This simple example might help the reader understand why *LR* is a quantity of importance when one faces the problem of discriminating between two populations.

More formal mathematical justifications are available for the use of *LR* for assessing the added value provided by new information $x$ when faced with discriminating between two situations. These justifications are based on ideal applications where the needed probabilities are exactly known. We give a brief outline of two theoretical justifications often given in the literature.

**Discriminating between two simple hypotheses**

Perhaps Neyman and Pearson are most recognized as the first to give a formal explanation for the role of the likelihood ratio in discriminating between two hypotheses, populations, or propositions. Suppose, in each of many repeated trials $T_i$ resulting in observations $x_i$ ($i = 1, \ldots, n$), one is tasked with deciding from which of two known distributions ($f_1$ or $f_2$) the observation $x_i$ is drawn. That is, in each trial one must decide between the hypotheses $H_{1i} : x_i$ came from $f_1$ or $H_{2i} : x_i$ came from $f_2$. The Neyman-Pearson fundamental lemma [A1] essentially states that these outcomes are optimally ordered according to the ratio

$$LR_i = \frac{f_1(x_i)}{f_2(x_i)},$$

in the sense that $x_i$ should be considered as more strongly favoring $H_{1i}$ than $x_{i'}$ favors $H_{1i'}$ if and only if $LR_i > LR_{i'}$. Given any rule $\mathcal{R}$ for discriminating between $H_{1i}$ and $H_{2i}$ that is based on an observation $x_i$ (*i.e.*, conclude $H_{1i}$ if $x_i$ satisfies some given condition and conclude $H_{2i}$ otherwise), one can always find an *LR* rule $\mathcal{R}_{LR}$ (*i.e.*, for a given $\tau \geq 0$, conclude $H_{1i}$ if $LR_i \geq \tau$ and conclude $H_{2i}$ if $LR_i < \tau$) that will, in the long run, correctly decide $H_{1i}$ to be true, when it is in fact true, for at least as many trials as $\mathcal{R}$ will, and will wrongly decide $H_{1i}$ to be true, when it is in fact false, in no more trials than $\mathcal{R}$ will.

Note that we have assumed $f_1$ and $f_2$ to be completely known. That is, no modeling was necessary and no distribution was fitted to empirical data. Neyman-Pearson fundamental lemma is applicable primarily in such *ideal* situations. Real situations are more complex and optimality of *LR* based rules cannot be guaranteed.

*LR* **in a Bayesian framework.**

The Bayesian framework is based on the philosophical viewpoint that all probabilities are personal and quantify one's state of uncertainty regarding the truth of propositions. Given the problem of discriminating between $H_1$ and $H_2$ as above, one first quantifies one's uncertainties associated with the truth of $H_1$ and of $H_2$ by (prior) probabilities $\pi_1$ and $\pi_2 = 1 - \pi_1$. These describe the levels of uncertainty experienced by an individual prior to seeing the data $x$. After seeing $x$, one is interested in the posterior probabilities $P(H_1|x)$ and $P(H_2|x)$ (note that $P(H_1|x) + P(H_2|x) = 1$) or, equivalently, the posterior odds

$$\text{Posterior Odds} = \frac{P(H_1|x)}{P(H_2|x)} = \frac{P(H_1|x)}{1 - P(H_1|x)}.$$

An application of Bayes rule for updating one's prior personal probabilities after having observed new information leads to the equation

$$\text{Posterior Odds for } H_1 = \frac{P(H_1|x)}{P(H_2|x)} = \frac{P(x|H_1)}{P(x|H_2)} \times \frac{P(H_1)}{P(H_2)} = \frac{f_1(x)}{f_2(x)} \times \text{ Prior Odds for } H_1.$$

If we define *weight of evidence* associated with $x$ for a particular individual to be the ratio of posterior odds (given $x$) of that individual to his or her prior odds (before observing $x$) then the above equation implies that $LR = \frac{f_1(x)}{f_2(x)}$ is to be viewed as the weight of the evidence provided by $x$ for $H_1$ for the individual making the probability assessments.

### Surrogate *LR*s as Discriminant Scores

Neyman-Pearson fundamental lemma tells us that the theoretical *LR* is the best summary of the information in $x$ for discriminating between $H_1$ and $H_2$. In this sense, we can say that, when $f_1$ and $f_2$ are known, *LR* is the best *discriminant score*. When $f_1$ and $f_2$ are not known, it is customary to use empirical information to find surrogates for $f_1$ and $f_2$ (i.e. models) and use these to construct a surrogate *LR* corresponding to an observed value $x$. Different models based on different sets of assumptions will lead to different *LR*s. These can all be helpful, some more than others, in discriminating between $H_1$ and $H_2$. We continue to refer to these surrogate *LR* values as discriminant scores. The performance characteristics of competing discriminant scores may be evaluated empirically using suitable, ground-truth known data through the use of receiver operating characteristic (ROC) plots. For a detailed discussion of ROC plots the reader is referred to Peterson & Birdsall [A2] and Green & Swets ([A3-A4]).

### Summary

The study of *LR* in theoretical settings provide useful guidance when dealing with problems of discriminating between two or more populations in real life applications. Since we never really know $f_1$ or $f_2$, however, we have to rely on available data and statistical models to develop surrogates for the theoretical *LR*s and no theoretical optimality properties may be claimed in the Neyman-Pearson setting. Even under the Bayesian framework there is no unique *LR*. A main thrust of the paper is to bring to the attention of the community that these surrogate *LR*s can have substantial disagreements with one another and no unique authoritative model from which to derive an *LR* for public consumption exists. The usefulness of any particular surrogate *LR* (sometimes referred to as an *LR* system; see Leegwater *et al.*, [A5]) has to be demonstrated empirically using tools such as ROC plots.

## Appendix A: References

[A1] Neyman J., Pearson E. S. (1933). On the problem of the most efficient tests of statistical hypotheses, *Philosophical Transactions of the Royal Society of London* A 231: 289–337.

[A2] Peterson, W. W. and Birdsall, T. G. (1953). The Theory of Signal Detectability - Part I. General Theory, *Technical Report No. 13, Electronic Defense Group*, Department of Electrical Engineering, Engineering Research Institute, University of Michigan, Ann Arbor.

[A3] Green, D. M. and Swets, J. A. (1966). *Signal detection theory and psychophysics*, Wiley.

[A4] Green, D. M. and Swets, J. A. (1974). *Signal detection theory and psychophysics* (a reprint with corrections of the original 1966 edition), Huntington, NY: Robert E. Krieger Publishing Co.

[A5] Leegwater, A. J., Didier, M., Sjerps, M., Vergeer, P. and Alberink, I. (2017), Performance Study of a Score-based Likelihood Ratio System for Forensic Fingermark Comparison, Journal of Forensic Sciences, doi: 10.1111/1556-4029.13339, Available online at: `onlinelibrary.wiley.com`

## Appendix B: Additional Results from the Glass Example

In this section of the appendix we display results for additional choices of *F* and $G_0$. Choices considered here for $G_0$ are explained below:

| | |
|---|---|
| Figure 10 | $\chi^2$ distribution with 3 df |
| Figure 11 | Example symmetric, unimodal, nonparametric distribution |
| Figure 12 | Example unimodal, nonparametric distribution |

The top plot in each figure shows the 95% Kolmogorov-Smirnov confidence band for the *CDF* of refractive indices from 49 fragments from a single window (Bennett Data). The empirical *CDF* is shown in gray. The faded red and green smooth curves, respectively, correspond to members of the chosen scale family with the smallest and largest scaling factors such that, the discrete distributions obtained by accounting for interval-censoring in the reported data (shown using solid red and solid green line segments, respectively), are entirely contained within the confidence band.

The bottom plot in each figure displays the *LR* values corresponding to various choices for *F*, reflected by position along the x-axis, and the scale factor used with the shape chosen for $G_0$. The left-most results correspond to the estimate of *F* labeled as Jump, which is displayed in Figure 2. The remaining positions reflect the bandwidth of the Gaussian kernel leading to the estimate of *F* used in computing the *LR*. Within each choice of *F*, the *LR* values are staggered in order of the scale parameter used with $G_0$ to emphasize the potential non-monotonic relationship between *LR* and the scale parameter. The points are color-coded to indicate the associated scale parameter values in accordance with the legend titled $\sigma_{\text{within}}$.

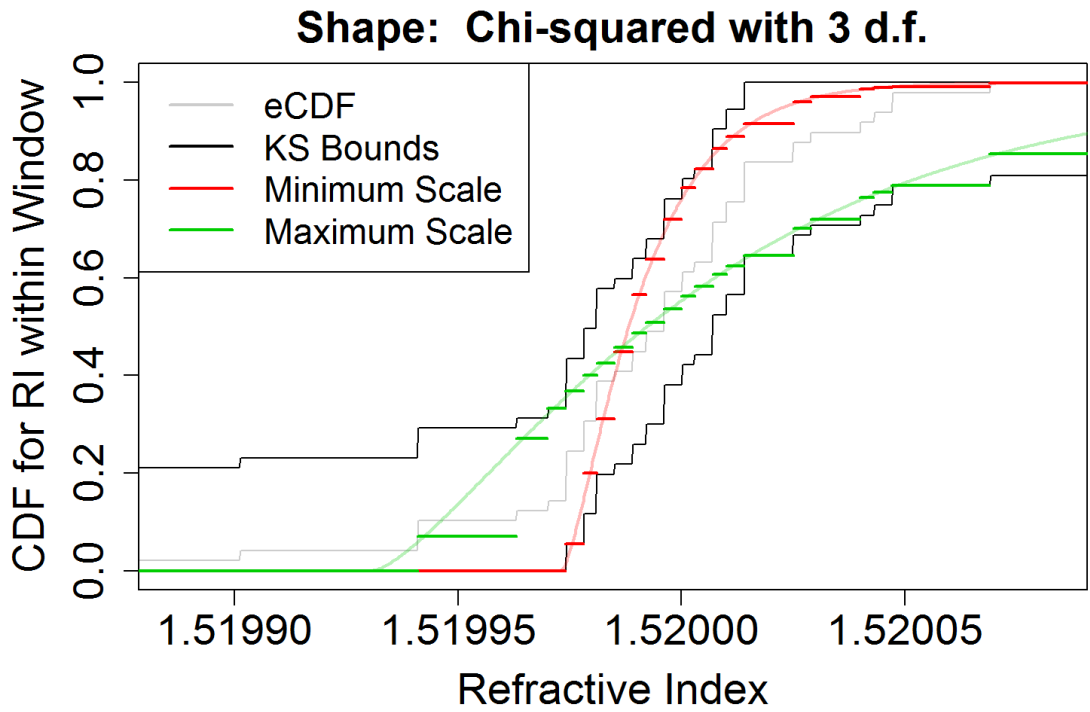

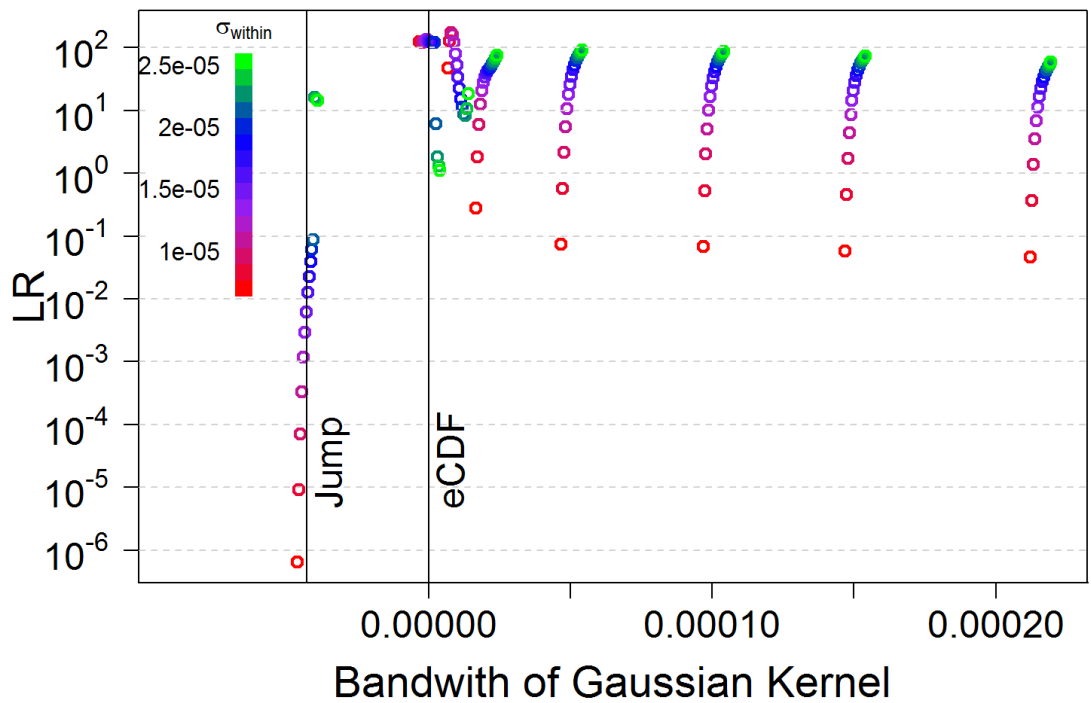

**Figure 10.** *LR* values when $G_0$ is a $\chi^2$ distribution with 3 degrees of freedom.

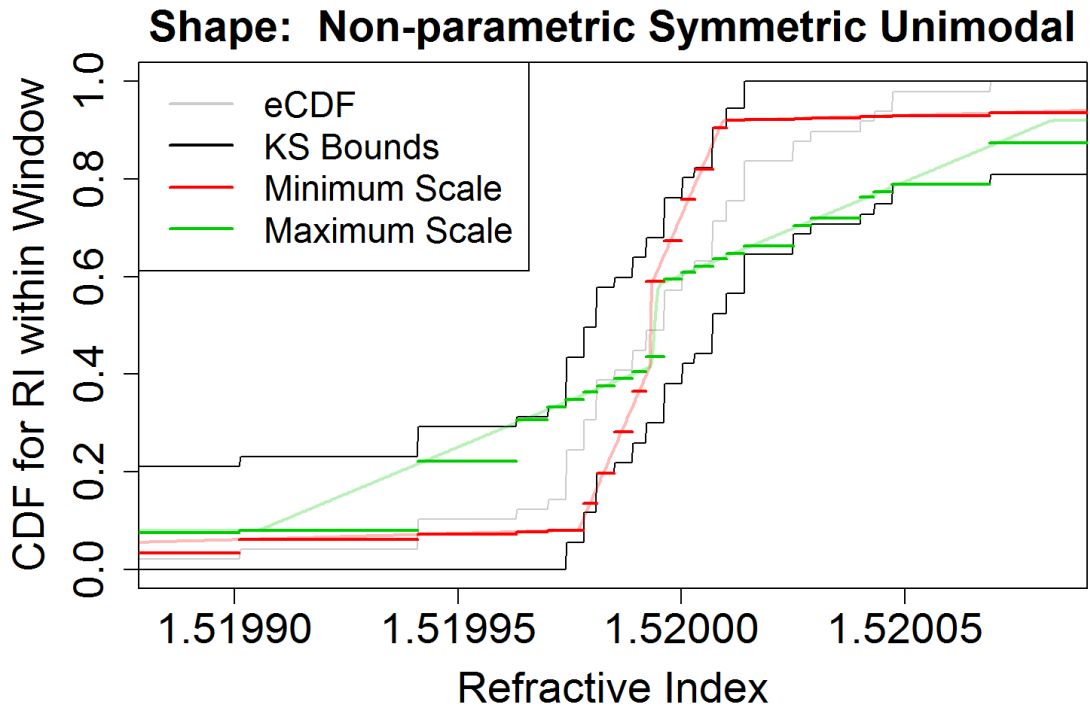

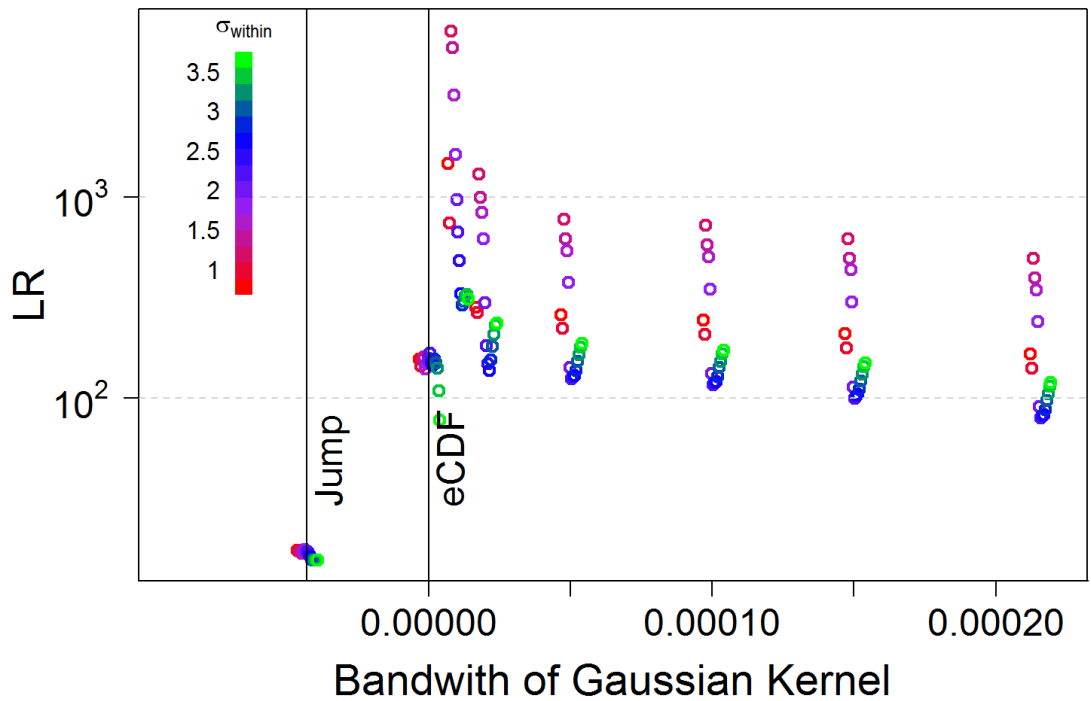

**Figure 11.** *LR* values when $G_0$ has the shape of the nonparametric symmetric unimodal distribution shown.

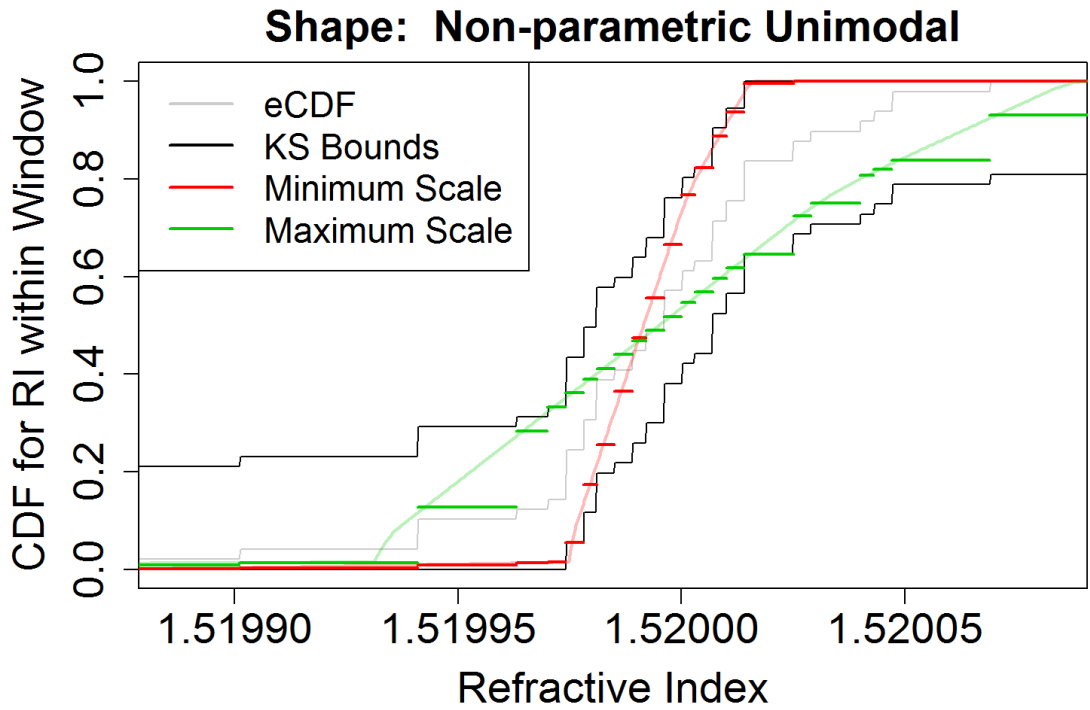
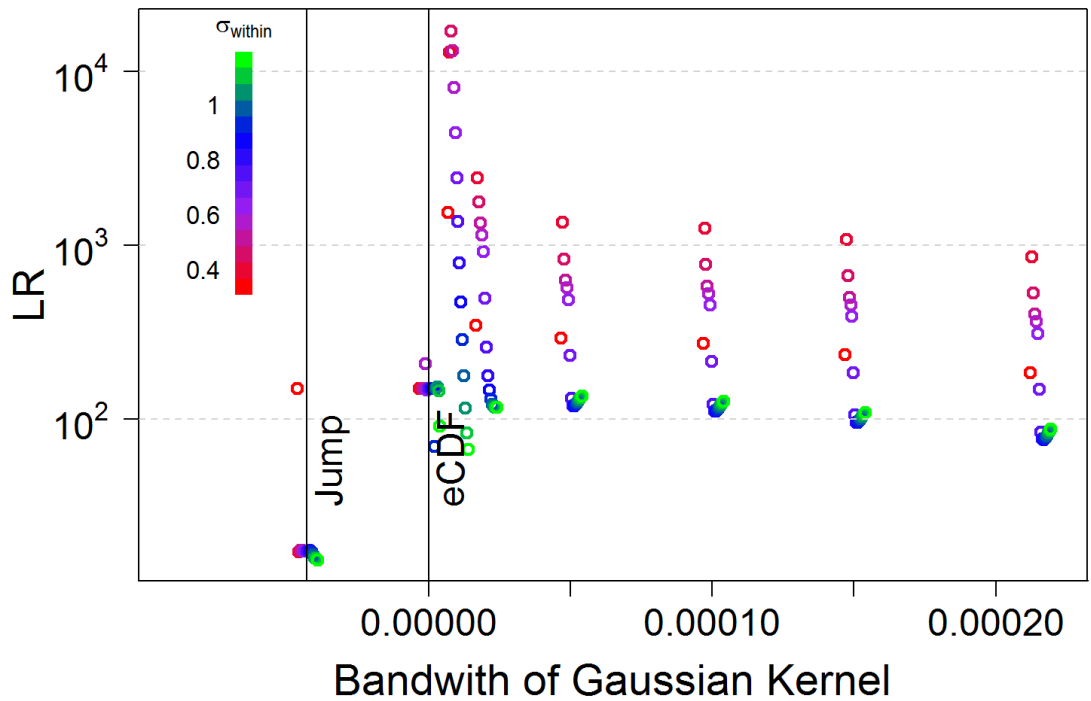

Figure 12. *LR* values when $G_0$ has the shape of the nonparametric unimodal distribution shown.